\documentclass[twocolumn]{aastex631} 

\shortauthors{S. Andreon et al.}
\shorttitle{Compton Y - mass relation of a gravity selected sample}
\graphicspath{{./}{figures/}}

\begin{document}

\title{Gravity-selected cluster samples: a new take on the Compton Y-mass relation}

\author{S. Andreon}
\affiliation{
INAF--Osservatorio Astronomico di Brera, via Brera 28, 20121, Milano, Italy\\}
\author{M. Radovich}
\affiliation{INAF--Osservatorio Astronomico di Padova, Vicolo Osservatorio 5, 35122, Padova, Italy\\}

\begin{abstract}

We investigate the scaling relation between the Compton Y parameter and mass in a gravity-selected sample of galaxy clusters, selected based on their gravitational lensing effects on background galaxies. Unlike ICM-selected samples, gravity-selected clusters do not require corrections for selection biases in intracluster medium (ICM) properties, such as Compton Y, since these properties are not used for selection.
Our sample consists of 13 gravity-selected clusters with weak-lensing signal-to-noise ratios greater than 7 and redshifts in the range $0.1 < z < 0.4$, drawn from the peak catalog of shear-selected clusters in Oguri et al. (2021). We determined spectroscopic redshifts using SDSS spectroscopy, derived tangential radial profiles from HSC shear data, and calculated cluster masses by fitting a Navarro, Frenk, \& White (1997) profile. 
Compton Y measurements were obtained from ACT maps.
Our sample reveals a broader scatter in the Compton Y-mass scaling relation compared to ICM-selected samples, even after accounting for selection effects and mass bias of the latter, that we find to be  $(1-b) = 0.72 \pm 0.09$.
Additionally, we identify a second population of clusters with unusually small Compton Y for their mass, which is absent in ICM-selected samples. This second population comprises $13^{+10}_{-8}$\% of the entire sample, with O19 currently being the only secure representative.
\end{abstract}


\section{Introduction} 

Most of our current knowledge about the properties of galaxy clusters is based on the
analysis of samples selected by their minority components, either galaxies or the intracluster medium (ICM hereafter).
These components account for about, or less than, 10\% of the cluster mass (e.g., Zwicky 1937, Vikhlinin et
al. 2006, Andreon 2010). Even after accounting for selection effects, the way the sample is selected affects
the inferred cluster properties, including: scatter around the X-ray luminosity-mass relation (Andreon et
al. 2011, 2016, 2025), the Temperature-mass relation (Andreon et al. 2022), the Compton Y-mass relation (Andreon et al. 2019, 2025), the gas fraction (Andreon et al. 2017a, Bigwood et al. 2024, Hadzhiyska et al. 2024), the X-ray core radius distribution (Andreon et al. 2024), the X-ray
morphological composition (Ecker et al. 2011, Rossetti et al. 2017), the radial electron and pressure
profiles (Andreon et al. 2019, 2021, 2022, 2023, 2025, Dicker et al. 2020, Di Mascolo et al. 2020, Sayers et al. 2023) and, for one cluster so far, the richness-mass relation (Andreon et al. 2025).

Besides incidental errors, which are always possible, the task of correcting for
selection effects is formidable because two terms are needed (Andreon \& Hurn 2013): the probability of
detection given the object parameters, $p(det|\theta)$, and the distribution of the object parameters,
$p(\theta)$ (e.g., how many more faint objects exist at a fixed mass per bright object).
The former can be computed in different ways, often by injecting simulated objects with
known values of the parameters in real
data and indeed was already considered in early computations
(e.g. Gioia et al. 1990, and Henry et al. 1992). The latter, i.e. the distribution of
objects in the parameter space, $p(\theta)$, is usually not
accessible when the studied objects are selected from survey data
because every survey has a minimal
threshold below which a large number of clusters might be. 
Therefore, it is unsurprising that in a Sunyaev-Zeldovich-selected sample, the  Compton Y vs mass scaling relation
is sensitive to the specifics of the adopted modeling (Shirasaki et al. 2024).

Selecting independently of the quantity being studied at fixed mass would be preferable
(e.g., Andreon et al. 2004, 2006, 2008, 2010, 2016). 
Selecting clusters by gravity is possible, because
the gravitational potential distorts the shape of background galaxies. The amount of distortion is proportional to mass, and therefore clusters can be detected, and their mass estimated, based on the deformation of background galaxy shapes, referred to as shear (Bartelmann \& Schneider 2001). By definition, samples selected by the gravitational potential alone (hereafter referred to as gravity-selected samples) are not expected to be biased against, or toward, objects with deviations above or below the mean relation with mass (e.g., with larger/smaller Compton Y) at fixed mass.
In fact, while in principle cluster elongation along the line of sight enhances the shear signal and, if shared by the intracluster medium (ICM), also boosts the Compton Y signal, elongation typically affects mass by less than 20\% (Gruen et al. 2015; Chen et al. 2020) and has an even smaller impact on Compton Y, as ICM elongation is generally a fraction, approximately half, of the total mass elongation (e.g., Suto et al. 2017; Shin et al. 2022). Consequently, the impact of selection effects induced by elongation is negligible for samples of the size analyzed in this study and are further mitigated by the partial alignment of elongation along the Compton Y-mass relation, reducing its overall impact.

Gravity selected samples have the advantage of not requiring corrections for selection biases in the ICM property being investigated, such as Compton Y, unlike studies of the ICM based on ICM-selected samples, where that property is used for selection. These gravity-selected samples are beginning to be utilized in cosmological investigations (e.g., Chiu et al. 2024).
On the other hand, the signal allowing cluster detection is weak and therefore subject to Eddington bias (Eddington 1913), which needs to be accounted for: most clusters with observed mass $M_{obs}$ have lower intrinsic mass $M$ due to the steepness of the mass function (Andreon et al. 2009, Andreon \& Congdon 2014, Andreon \& Weaver 2014).

A pilot study based on just four gravity-selected clusters by Andreon et al. (2025) finds ICM properties that are rare at best in ICM-selected samples, and even more so in samples as small as those considered by these authors, clearly indicating that our knowledge of cluster properties based on ICM-selected samples is biased, even after accounting for the selection biases of ICM-selected samples. Compton Y is one of the ICM properties considered by the authors, who found two out of four clusters in a region of the Compton Y-mass plane not populated by ICM-selected samples.

In this work, we aim to investigate the Compton Y vs. mass scaling more thoroughly with a larger sample, enabling us to derive the parameters of the relation between these quantities.

Throughout this paper, we assume $\Omega_M=0.3$, $\Omega_\Lambda=0.7$, and $H_0=70$ km s$^{-1}$ Mpc$^{-1}$. 
Gaussian posteriors are summarized
in the form $x\pm y$, where $x$ and $y$ are 
the mean and standard deviation. 
Non Gaussian posteriors are summarized as $x^{+y}_{-z}$ where $(x-z),x$, and $x+y$ are the
$(16,50,84)$ percentiles. All logarithms are in base 10.

\begin{figure}
\centerline{\includegraphics[width=9truecm]{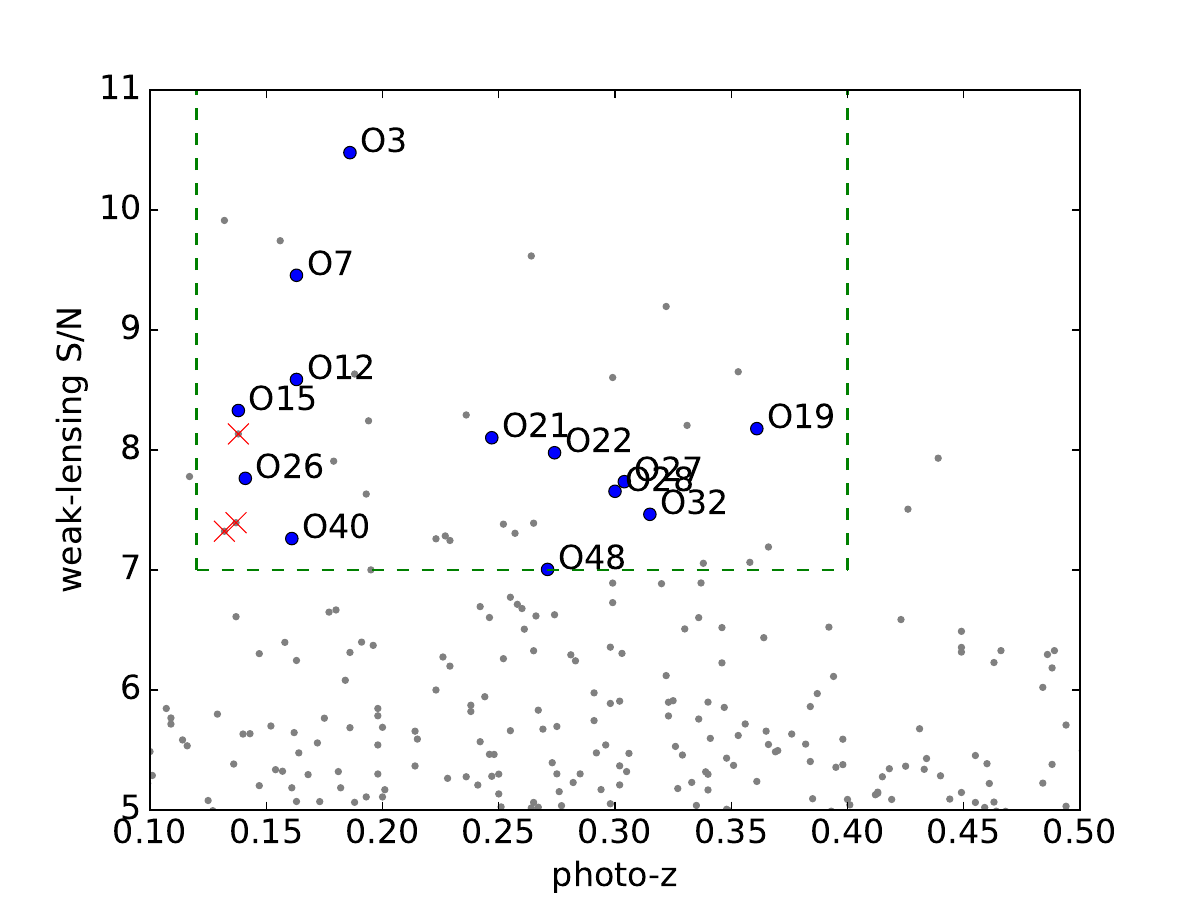}}
\caption{Weak-lensing detection S/N vs cluster photometric redshift. Gray points are detections below $\delta=20$ deg, approximatively the northern limit of the ACT map.
The green dashed contour delimits
the considered signal-to-noise and redshift ranges.  Blue points  
inside this area are inside the HSC
DR1 shear and ACT footprints and therefore form our sample. 
The crosses indicate peaks removed from the sample because part of a complex system. 
}
\label{fig:SNz}
\end{figure}

\section{Sample selection, weak-lensing masses and Compton Y}

\subsection{Sample selection and shear profiles of the gravity-selected sample}

The sample of gravity-selected clusters is composed by TI05 weak-lensing peaks in Oguri et al. (2021) with signal-to-noise ratios $S/N > 7.001$ and photometric redshifts $0.12 < z_{phot} < 0.4$, located within the overlapping footprints of the ACT Compton $Y$ maps (Coulton et al. 2023) and the HSC DR1 shape catalog (Mandelbaum et al. 2018). The peaks satisfying these criteria are 16 (Fig.~\ref{fig:SNz}). The Oguri et al. (2021) shear peak catalog was produced using very deep observations with excellent seeing over approximately 500 deg$^2$ of the sky, observed by the Hyper Suprime-Cam (HSC) as part of a Subaru Strategic Program (Aihara et al. 2022). We restricted our sample to $S/N \gtrsim 7$ to limit the magnitude and uncertainty of the Eddington (1913) correction.

\begin{figure}
\centerline{\includegraphics[width=9truecm]{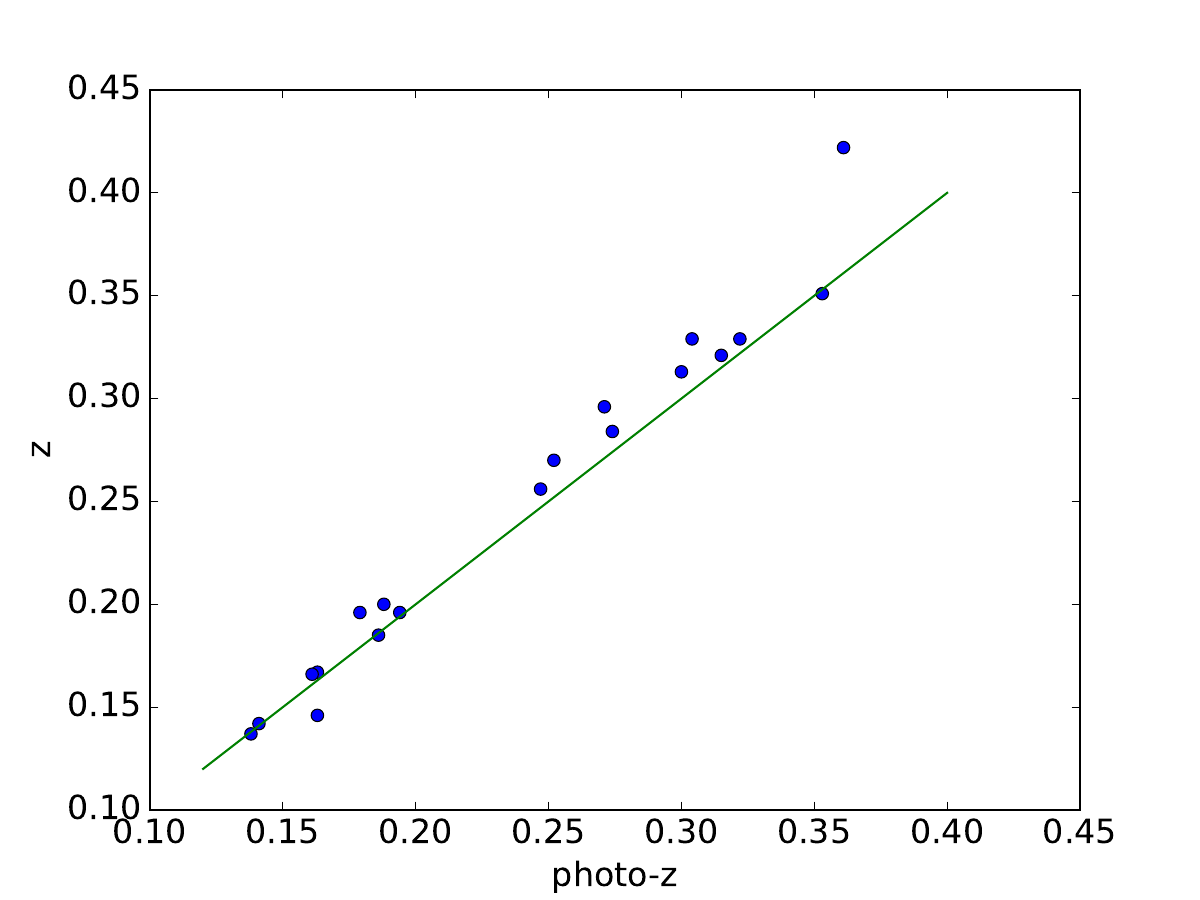}}
\caption{Photometric vs spectroscopic redshift of the studied sample and a few more clusters inside the ACT footprint but outside the HSC DR1 shape catalog. The green line indicates the one-to-one relation. The most distant cluster is also the one with larger $|z_{\rm phot}-z_{\rm spec}|$.}
\label{fig:zz}
\end{figure}

Three peaks (O20, O34, and O37) correspond to a complex system (Abell 1881) consisting of three interacting clumps in the process of merging, two of which are separated by approximately $r_{200}$. All three optical clumps exhibit corresponding weak-lensing and SZ detections and share the same spectroscopic redshift. However, it is unclear whether they constitute a single system or multiple systems. Furthermore, the weak-lensing analysis of this system cannot assume spherical symmetry at $r_{200}$. We therefore exclude this complex system from our sample. Since the exclusion is unrelated to the Compton Y signal, the sample
continues to be gravity-selected and does not need to account for a never-applied Compton Y selection.

As a result, the final sample consists of 13 gravity-selected clusters with $S/N > 7.001$, $0.12 < z_{phot} < 0.4$, located within the overlap of the ACT and shear catalog footprints. These clusters are shown as blue points in Fig.\ref{fig:SNz} and listed in Tab.\ref{tab1}.

Using SDSS DR18 spectroscopic data (Almeida et al. 2023), we measured the spectroscopic redshift of all targets (Fig.~\ref{fig:zz}), whose  photometric redshift is given in  Oguri et al. (2021). 
As cluster redshift, we considered the median of the main peak of the redshifts distribution of the galaxies within 1 Mpc from the cluster center. As cluster center we adopt the value reported
in Oguri et al. (2021) and we checked that it matches the Compton Y peak. 
The most distant cluster, O19, has a spectroscopic redshift of $z_{\rm spec}=0.422$, scattering from its photometric redshift of $z_{phot}=0.361$.

\begin{figure*}
\centerline{\includegraphics[width=6truecm]{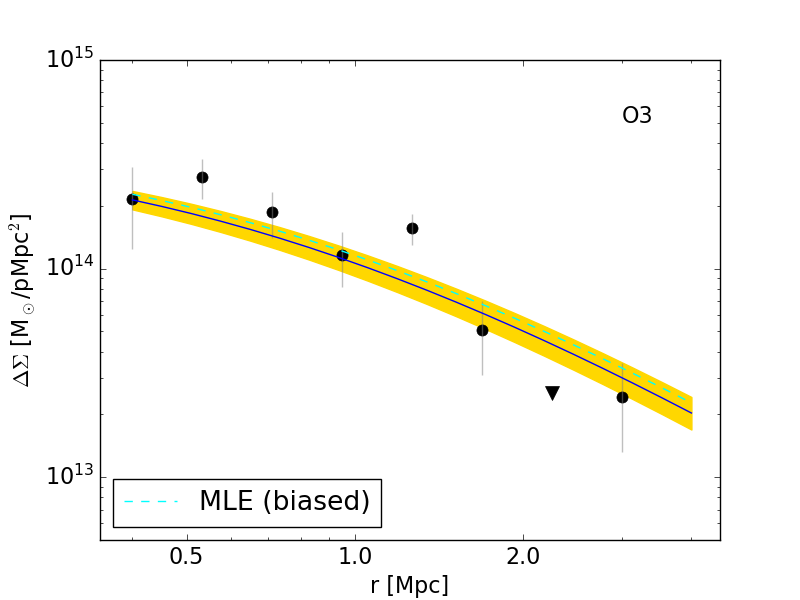} \includegraphics[width=6truecm]{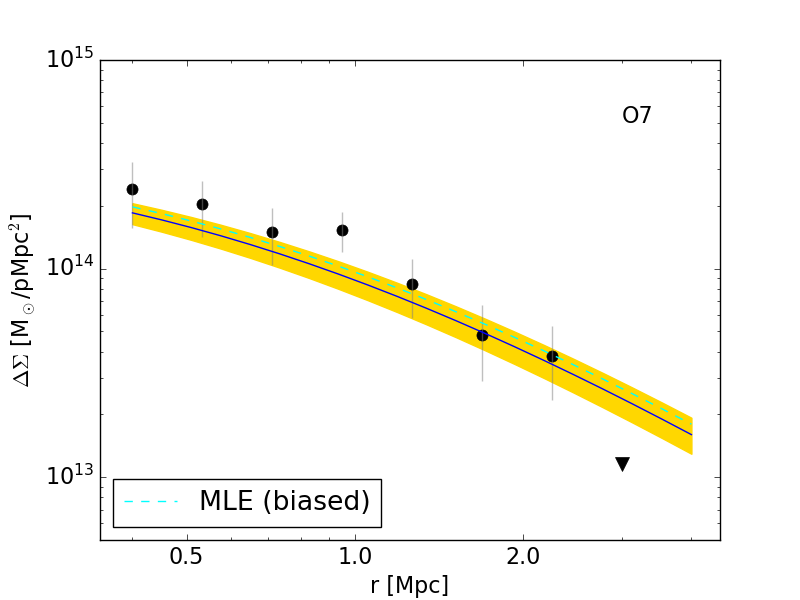}\includegraphics[width=6truecm]{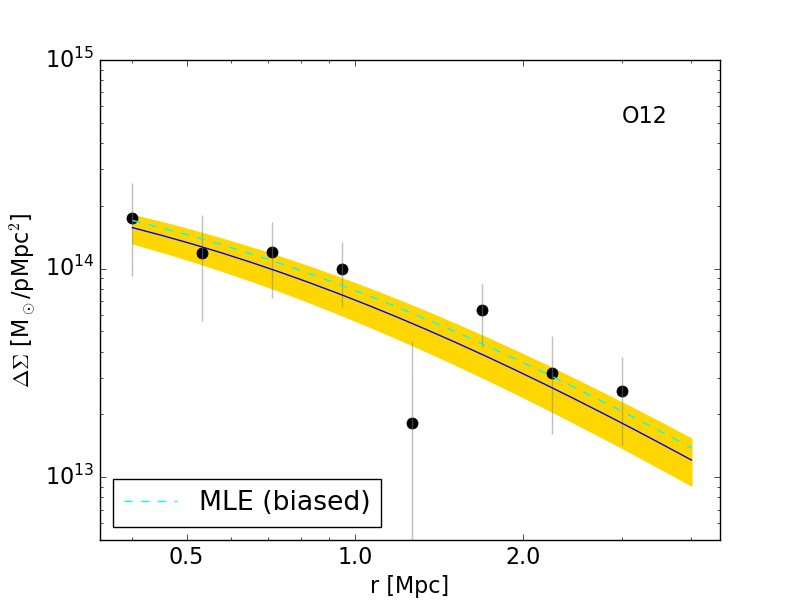}}
\centerline{\includegraphics[width=6truecm]{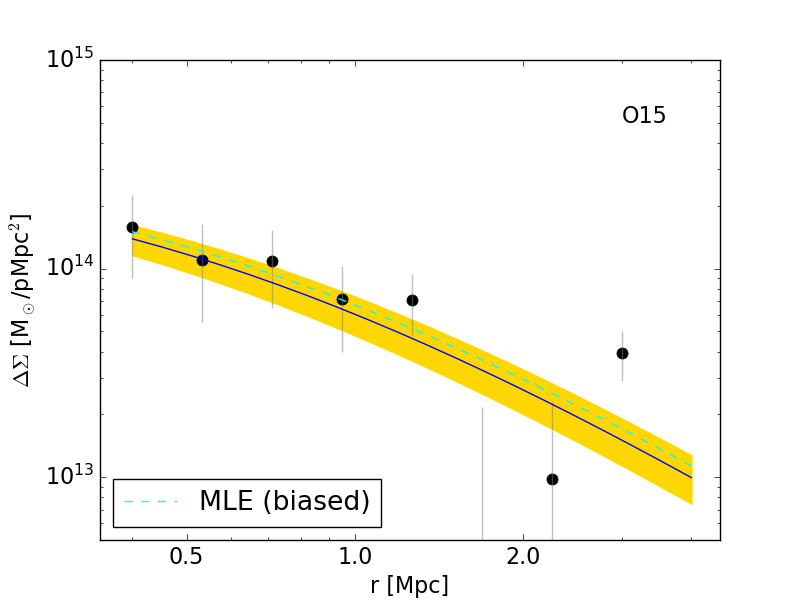} \includegraphics[width=6truecm]{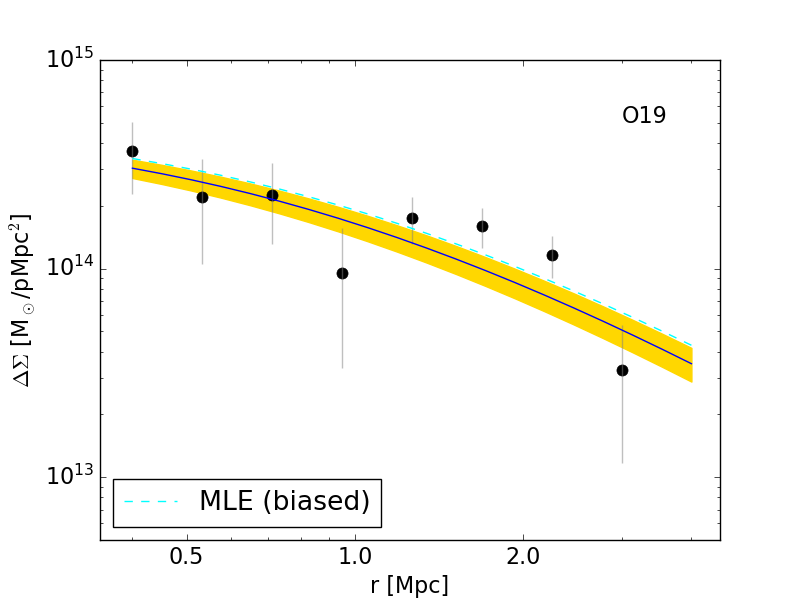}\includegraphics[width=6truecm]{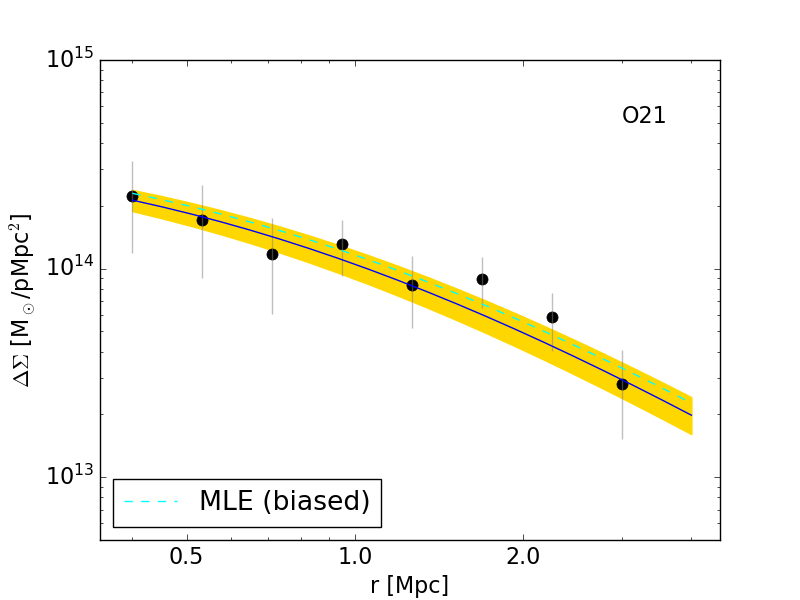}}
\centerline{\includegraphics[width=6truecm]{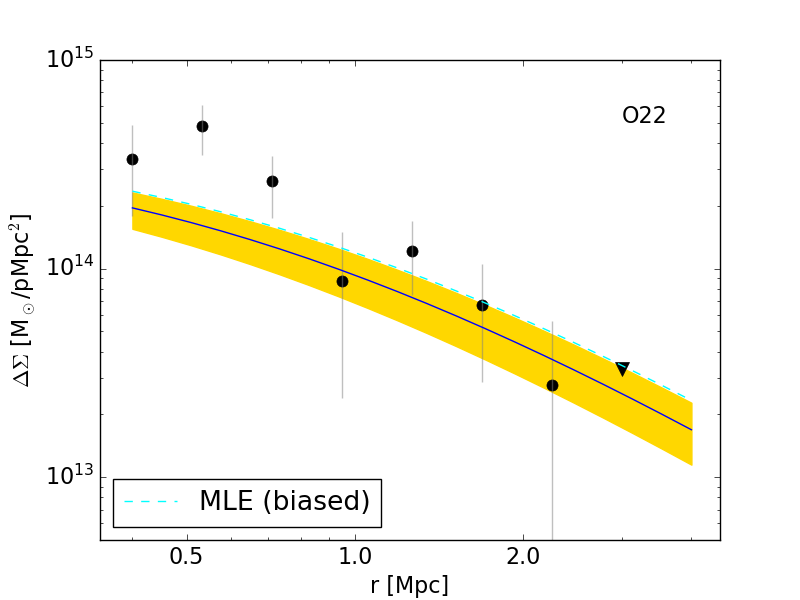} \includegraphics[width=6truecm]{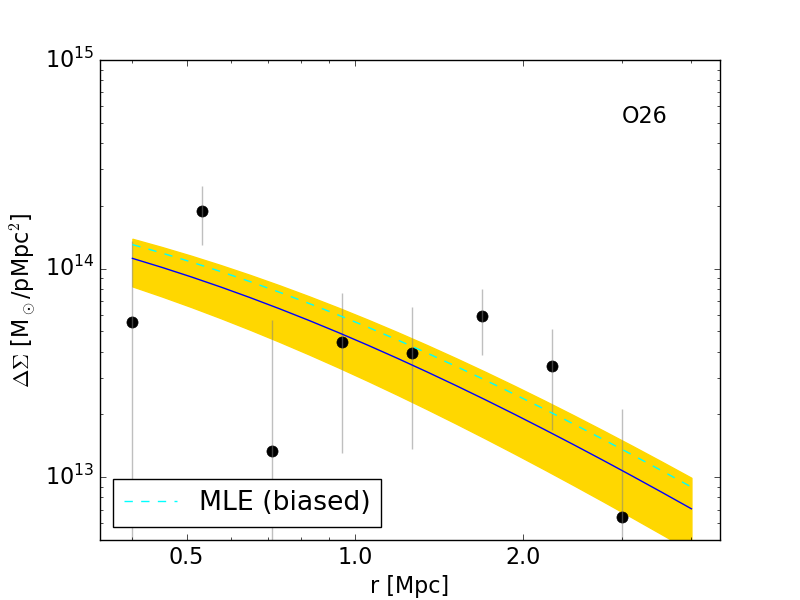}\includegraphics[width=6truecm]{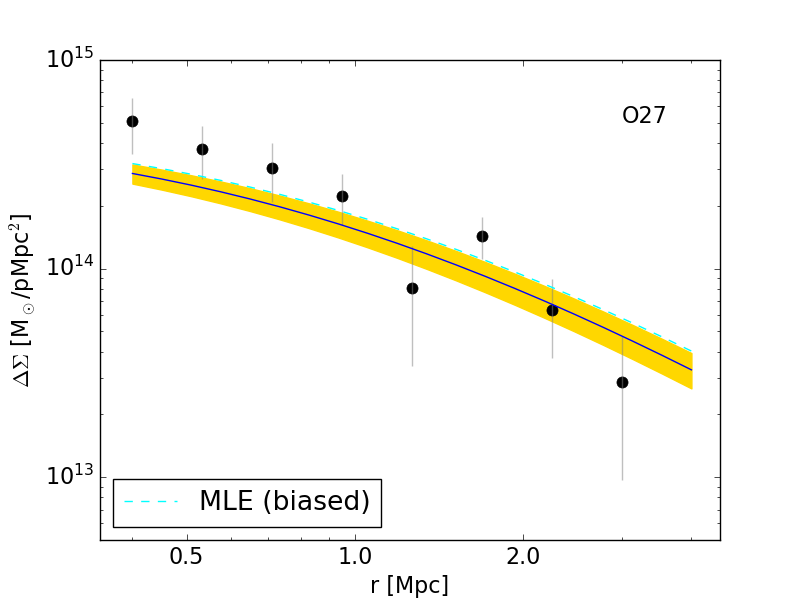}}
\centerline{\includegraphics[width=6truecm]{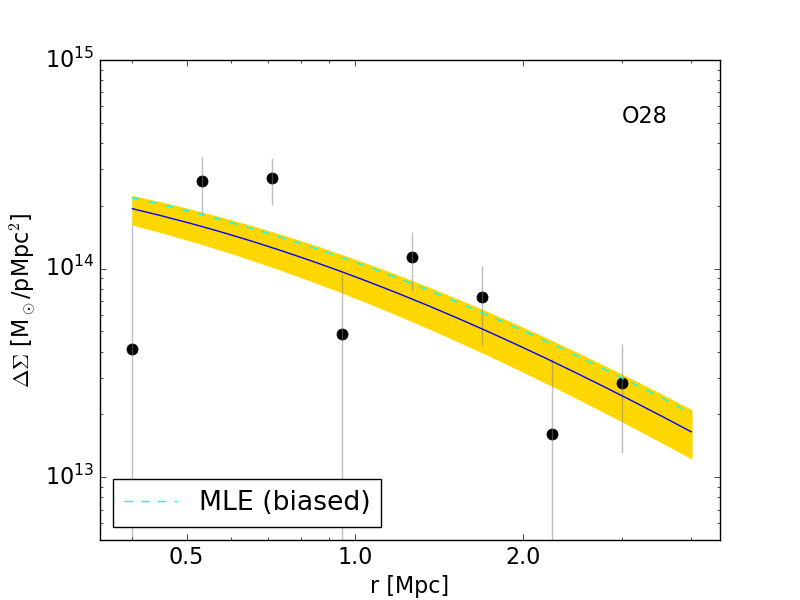} \includegraphics[width=6truecm]{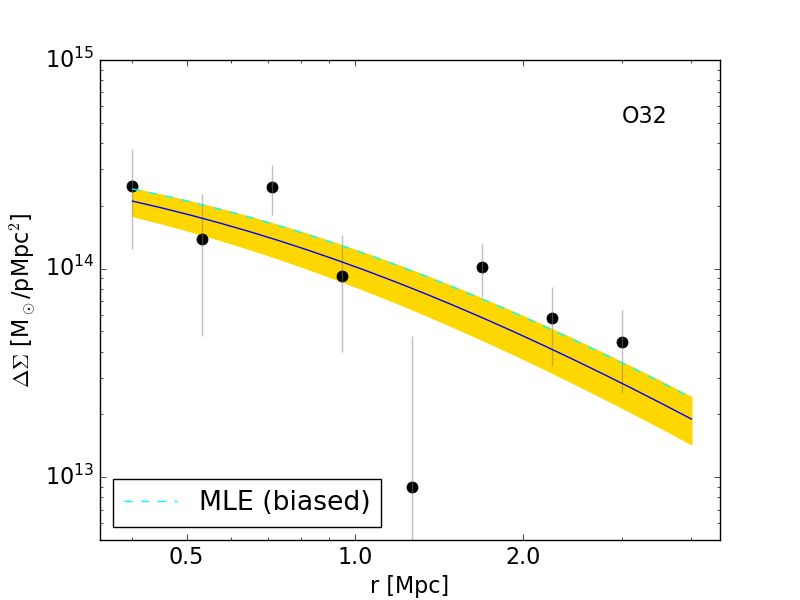}\includegraphics[width=6truecm]{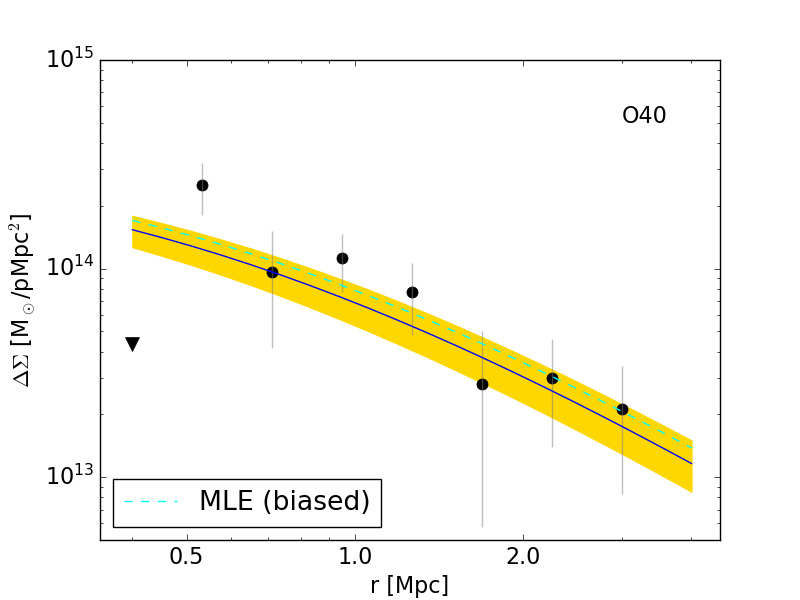}}
\centerline{\includegraphics[width=6truecm]{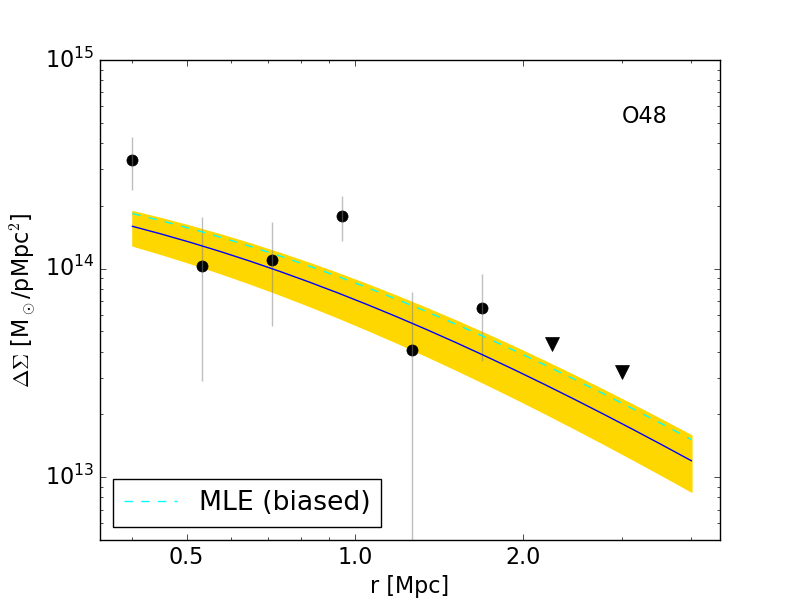} }
\caption{Binned tangential shear profile of the gravity-selected clusters. The solid line with yellow shading indicates the mean model and 68\% uncertainty. Uncertainty in the model also accounts for intrinsic scatter, whereas the plotted error bars only account for shape noise and large scale structure.  The dashed cyan line shows the maximum likelihood estimate (MLE), which is biased (by a minimal amount, given our choice of the S/N). The triangles are $2\sigma$ upper limits. }
\label{fig:WL}
\end{figure*}

\begin{figure*}
\centerline{\includegraphics[width=2.5truecm]{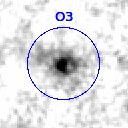} \includegraphics[width=2.5truecm]{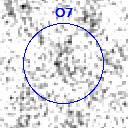}\includegraphics[width=2.5truecm]{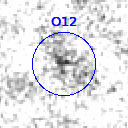}
\includegraphics[width=2.5truecm]{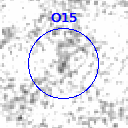} \includegraphics[width=2.5truecm]{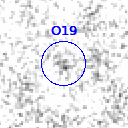}
\includegraphics[width=2.5truecm]{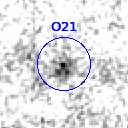}
\includegraphics[width=2.5truecm]{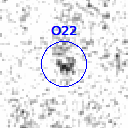}} \centerline{\includegraphics[width=2.5truecm]{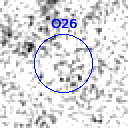}\includegraphics[width=2.5truecm]{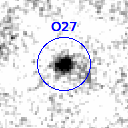}
\includegraphics[width=2.5truecm]{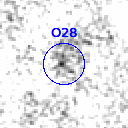} \includegraphics[width=2.5truecm]{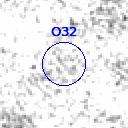}\includegraphics[width=2.5truecm]{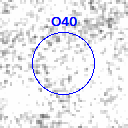}
\includegraphics[width=2.5truecm]{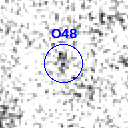}\includegraphics[width=2.5truecm]{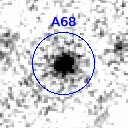} }
\centerline{\includegraphics[width=2.5truecm]{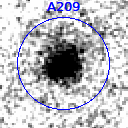}\includegraphics[width=2.5truecm]{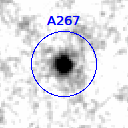}
\includegraphics[width=2.5truecm]{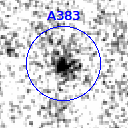} \includegraphics[width=2.5truecm]{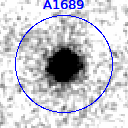}\includegraphics[width=2.5truecm]{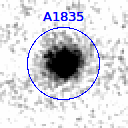}
\includegraphics[width=2.5truecm]{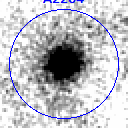}\includegraphics[width=2.5truecm]{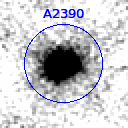} }
\caption{Compton-y maps of the gravity-selected (O* labels, first 13 panels) and X-ray selected (A* labels, last 8 panels; the cluster with truncated label is Abell 2204) samples. The circle marks $r_{200}$. All images are displayed with identical intensity ranges. Each map is 32x32 arcmin wide. North is up and East is to the left.}
\label{fig:Ymaps}
\end{figure*}

Tangential shear profile derivation and analysis follow the methodology described in Andreon et al. (2024), using the same code and shear catalog (of a different cluster sample, with one cluster (O21) in common with the present study). Briefly, we use the first-year shape catalog and the galaxy selection criteria outlined in Mandelbaum et al. (2018).
The surface density of background galaxies after applying photometric and photometric-redshift cuts averages 12 galaxies per arcmin$^2$, depending on the cluster. We used radial bins with equal logarithmic spacing ($\Delta \log{r} = 0.125$) and computed the tangential shear profile as described in prior works \citep[see, e.g.,][]{Umetsu2020}, accounting for shape measurement uncertainty, rms ellipticity, calibration bias, and the responsivity factor \citep[e.g.,][]{Mandelbaum2018}.

The derived radial tangential shear profiles of the clusters are presented in Fig.~\ref{fig:WL}. Nine clusters (all except O3, O7, O26, and O48) lie within the footprint of the KiDS shear data Kuijken et. (2019), and we verified that the profiles derived from KiDS data, although much noisier, are consistent with those obtained from HSC data.

As in Andreon et al. (2024), we fit the tangential radial profiles using a Navarro et al. (1997) radial profile with a 
concentration-mass relation from Dutton \& Macci\`o (2014) 
and a uniform prior on $\log$ concentration between $\log{0.1}$ and $\log{30}$, 
following Hamana et al. (2023).  

The mass prior, derived from the Tinker et al. (2008) mass function, accounts for Eddington bias. In all fits, we account for galaxy shape noise and include a 20\% intrinsic scatter in the cluster lensing signal due to elongation, triaxiality, and presence of correlated halos, based on the results of Gruen et al. (2015) and Chen et al. (2020). To mitigate potential deviations from the weak-lensing regime, inaccuracies in shape measurements due to dense cluster environments, and the effects of other clusters or large-scale structures, we only use galaxies within the range 350 kpc to 3.5 Mpc.
We do not account for cosmic shear covariance from uncorrelated large-scale structures along the line of sight, as this contribution is negligible (Wu et al. 2019). Fig.~\ref{fig:WL} illustrates the fits to the tangential shear profiles of our clusters, along with the fits obtained when the Eddington correction is ignored. Given the high signal-to-noise ratio of the shear data, the Eddington correction is smaller than the statistical uncertainties.
Table~\ref{tab1} presents the derived masses of the clusters.

\subsection{An additional, L$_X$-selected, sample}

As an additional sample, we use eight X-ray-selected LoCUSS clusters (Smith et al. 2016) from Okabe \& Smith (2016) that meet the criteria of weak-lensing signal-to-noise $S/N > 7.001$ and $0.12 < z < 0.4$ within the ACT footprints.
The parent sample is purely X-ray-selected, based on cuts in ROSAT X-ray luminosities and redshifts. All of these clusters have weak-lensing masses reported in Okabe \& Smith (2016), which we use in this work. These masses are derived from deep Subaru Suprime-Cam (Miyazaki et al. 2002) images acquired under excellent seeing conditions.
Unlike us, the authors do not apply the Eddington (1913) correction. However, given the high signal-to-noise ratio of the data, this correction is negligible, as is the case of our sample (although we apply the correction in our analysis).

\subsection{Compton Y}

For all clusters in our sample, we measured the cylindrical Compton Y, $Y_{{\rm cyl},200}$, within apertures of $r_{200,{\rm wl}}$ radii using ACT Compton-y maps (Coulton et al. 2024). Fig.~\ref{fig:Ymaps} shows cutouts of the Compton-y map centered on the clusters. The data allow us to measure fluxes for all clusters except O40, for which only an upper limit could be determined. This cluster is retained in the sample to avoid introducing a selection bias.

Errors are estimated based on the scatter of measurements in apertures at random centers distributed around each cluster. The spherical Compton Y, $Y_{{\rm sph},200}$, is derived from $Y_{{\rm cyl},200}$ by dividing the latter by a factor of 1.05, which is appropriate for a universal pressure profile with Arnaud et al. (2010) parameters.

The flux derivation accounts for the rectangular projection used in the ACT Compton-y map by utilizing the {\rm pixell} package\footnote{https://github.com/simonsobs/pixell}. For our clusters, the aperture radius is significantly larger than the ACT beam, and therefore, point spread function effects are negligible.

Since both Compton Y and mass are measured within the same radius $r_{200}$, we expect a covariance between these quantities. We estimated this covariance by performing Compton Y measurements at different radii and found that the radial covariance is negligible compared to the associated errors. Consequently, the covariance is not included in our analysis.

Our Compton Y measurements are reported in Table~\ref{tab1}.

\begin{table}
\caption{Gravity-selected (O*) and X-ray selected (A*) cluster sample and results of our analysis.}
\scriptsize
\setlength{\tabcolsep}{4pt}
\hskip -1.3 truecm 
\begin{tabular}{lrrrrrrr}
\hline
ID & \multicolumn{1}{c}{RA} & Dec &  z$_{\mathrm spec}$ & logM$_{200}$ & err & logY$_{200}$ & err  \\
 & \multicolumn{2}{c}{J2000} & & [M$_\odot$] & dex & Mpc$^{2}$ & dex  \\
\hline
O3 &  37.9224 & -4.8789 & 0.185   & 14.81 & 0.09    & -4.33 & 0.09     \\
O7 &  336.0368 & 0.3293 & 0.146   & 14.68 & 0.11    & -5.08 & 0.23      \\
O12 &  180.4305 & -0.1903 & 0.167   & 14.53 & 0.14    & -4.64 & 0.09  	 \\
O15 &  177.5865 & -0.6026 & 0.137   & 14.43 & 0.14    & -5.07 & 0.19  	 \\
O19 &  130.5909 & 1.6474 & 0.422   & 15.06 & 0.10    & -5.05 & 0.40    \\
O21 &  179.0472 & -0.3361 & 0.256   & 14.78 & 0.11    & -4.42 & 0.08    \\
O22 &  139.7018 & 2.2062 & 0.284   & 14.68 & 0.19    & -4.77 & 0.17     \\
O26 &  336.2271 & -0.3654 & 0.142   & 14.24 & 0.22    & -5.31 & 0.23   \\
O27 &  139.0410 & -0.4010 & 0.329   & 15.04 & 0.11    & -4.25 & 0.08    \\
O28 &  223.0843 & -0.9708 & 0.313   & 14.67 & 0.14    & -4.51 & 0.12    \\
O32 &  217.6835 & 0.8127 & 0.321   & 14.74 & 0.14    & -5.04 & 0.37    \\
O40 &  138.5191 & 1.6512 & 0.166   & 14.51 & 0.16    & -4.96 & -   \\
O48 &  336.4214 & 1.0749 & 0.296   & 14.50 & 0.17    & -4.94 & 0.23    \\
\hline      
A68 &  9.2785 & 9.1566 & 0.255   & 14.98 & 0.08    & -4.05 & 0.06      \\
A209 &  22.9689 & -13.6112 & 0.206   & 15.26 & 0.07    & -3.75 & 0.03   \\
A267 &  28.1748 & 1.0072 & 0.230   & 14.93 & 0.08    & -4.18 & 0.07     \\
A383 &  42.0141 & -3.5291 & 0.188   & 14.87 & 0.10    & -4.35 & 0.10   \\
A1689 &  197.8730 & -1.3410 & 0.183   & 15.20 & 0.06    & -3.80 & 0.03  \\ 
A1835 &  210.2588 & 2.8786 & 0.253   & 15.16 & 0.07    & -3.72 & 0.03   \\
A2204 &  248.1956 & 5.5758 & 0.152   & 15.14 & 0.09    & -3.85 & 0.03  \\
A2390 &  328.4034 & 17.6955 & 0.233   & 15.18 & 0.07    & -3.64 & 0.02  \\
\hline      
\end{tabular} 
\hfill\break
IDs and coordinates of gravity-selected clusters are taken from Oguri et al. (2021). 
IDs, coordinates, redshift, masses and errors on mass of the
X-ray selected sample are taken from Okabe \& Smith (2016 and private communication). 
The tabulated logY$_{200}$ is the spherical Compton Y. The value reported for O40 is the 95\% upper limit.
\label{tab1}
\end{table}

\section{Y-M scaling relation}

We now perform three analyses, each involving different assumptions: we first inspect the location of our clusters in the Compton Y vs. mass plane compared to literature scaling relations. This empirical approach is nearly assumption-free. Next, we fit our gravity-selected sample using a hypothesis-parsimonious analysis. This method yields a robust Compton Y vs. mass scaling relation, albeit with large uncertainties. Finally, we attempt a more ambitious analysis that incorporates our X-ray-selected sample, providing strong constraints on the Compton Y scaling at the cost of relying on a strong assumption.

All data are plotted and analyzed at a fiducial redshift close to the mean redshift of the studied samples, $z=0.25$. Within the studied redshift range, we assume self-similar evolution, modeled as E(z)$^{-2/3}$. Keeping the data at a redshift near the sample average, at odd with past literature on this subject, avoids extrapolating the Compton Y intercept of the scaling relation to redshifts outside the range sampled by observations. This approach ensures that measurements remain at the redshifts where they were originally obtained.
This choice also has the advantage of making the intercept  of the scaling relation largely independent of the assumed slope for the E(z) evolution, which is commonly taken to be $-2/3$ in the literature, including Planck Collaboration XX (2014). However, Andreon (2014) found this assumption to be inconsistent with the data from Planck Collaboration XX (2014).

\begin{figure}
\centerline{\includegraphics[width=9truecm]{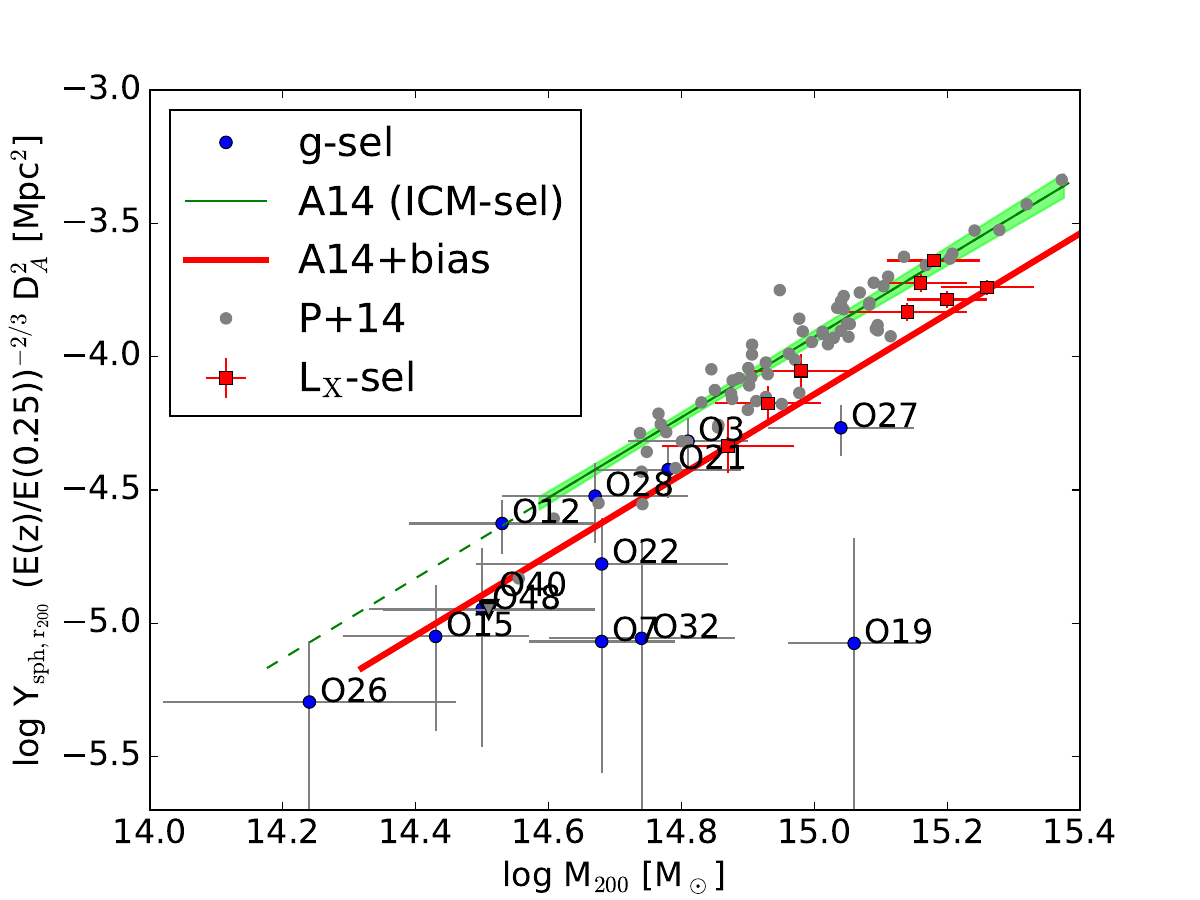}}
\caption{$Y_{{\rm sph},200}-M$ plot 
for various samples, as detailed in the inset. The downward pointing triangle at $\log M/M_\odot \sim 14.5$ indicates the $2\sigma$ upper limit of O40.
Gray points are clusters from Planck Collaboration XX (2014).
The solid green line is the relations found by Andreon (2014) using an ICM-selected sample. Points and scaling 
relation are (said to be) corrected for sample selection and
represent the true scaling of the whole population
of galaxy clusters. 
Gravity-selected clusters (blue points) are systematically below the ICM-selected sample (gray points)
indicating that the latter does not capture the variety of clusters in spite of the correction for sample selection.
Even after accounting for the mass bias (red line), the ICM-based scaling relation does not capture the variety of
clusters, as better visibile in Fig.~\ref{fig:Y_M_resid}.}
\label{fig:Y_M}
\end{figure}

\subsection{Empirical approach}

Fig.~\ref{fig:Y_M} shows the Compton Y vs. mass plot for the Planck Collaboration XX (2014) sample, which consists of 71 clusters (gray points) that are either X-ray or SZ-selected (i.e., ICM-selected). The masses of the Planck Collaboration XX (2014) sample are $Y_X$-based and Eddington (1913) corrected, while Compton Y is corrected by the Planck Collaboration XX (2014) for selection effects. Therefore, scaling relations derived from these 71 clusters represent the relationship between Compton Y and mass for the entire population of galaxy clusters, and indeed, it is used in this way by the Planck Collaboration XX (2014) paper.

Fig.~\ref{fig:Y_M} also shows the Compton Y vs. mass plot for our gravity-selected sample (blue points). Their position in the Compton Y vs. mass plane is unaffected by $Y$-related selection effects, simply because there is no $Y$-related selection for them. In other words, the probability of including any of the clusters (say O15) at fixed mass is the same, regardless of its $Y$ signal, because the sample selection does not filter clusters in or out based on large or small Compton Y at a fixed mass. These gravity-selected clusters can be directly compared to the 71 ICM-selected clusters meant to represent the whole population of galaxy clusters. They are systematically below the Planck data, in agreement with 
indications in Andreon et al. (2025).

The solid line with lime shading in Fig.~\ref{fig:Y_M} represents the Compton Y vs. mass scaling derived by Andreon (2014) using Planck data. Both data points and scaling relations were originally derived at overdensity $\Delta = 500$ and converted by us to $\Delta = 200$ assuming an NFW profile with the mean concentration corresponding to the mean mass of our sample (using Ragagnin et al. 2023) and a universal pressure profile (Arnaud et al. 2010), which was originally used by the Planck team to derive Compton Y values at $\Delta = 500$. The gravity-selected clusters
are systematically below the relations derived by Andreon (2014).

\begin{figure}
\centerline{\includegraphics[trim=0 0 0 100,clip,width=9truecm]{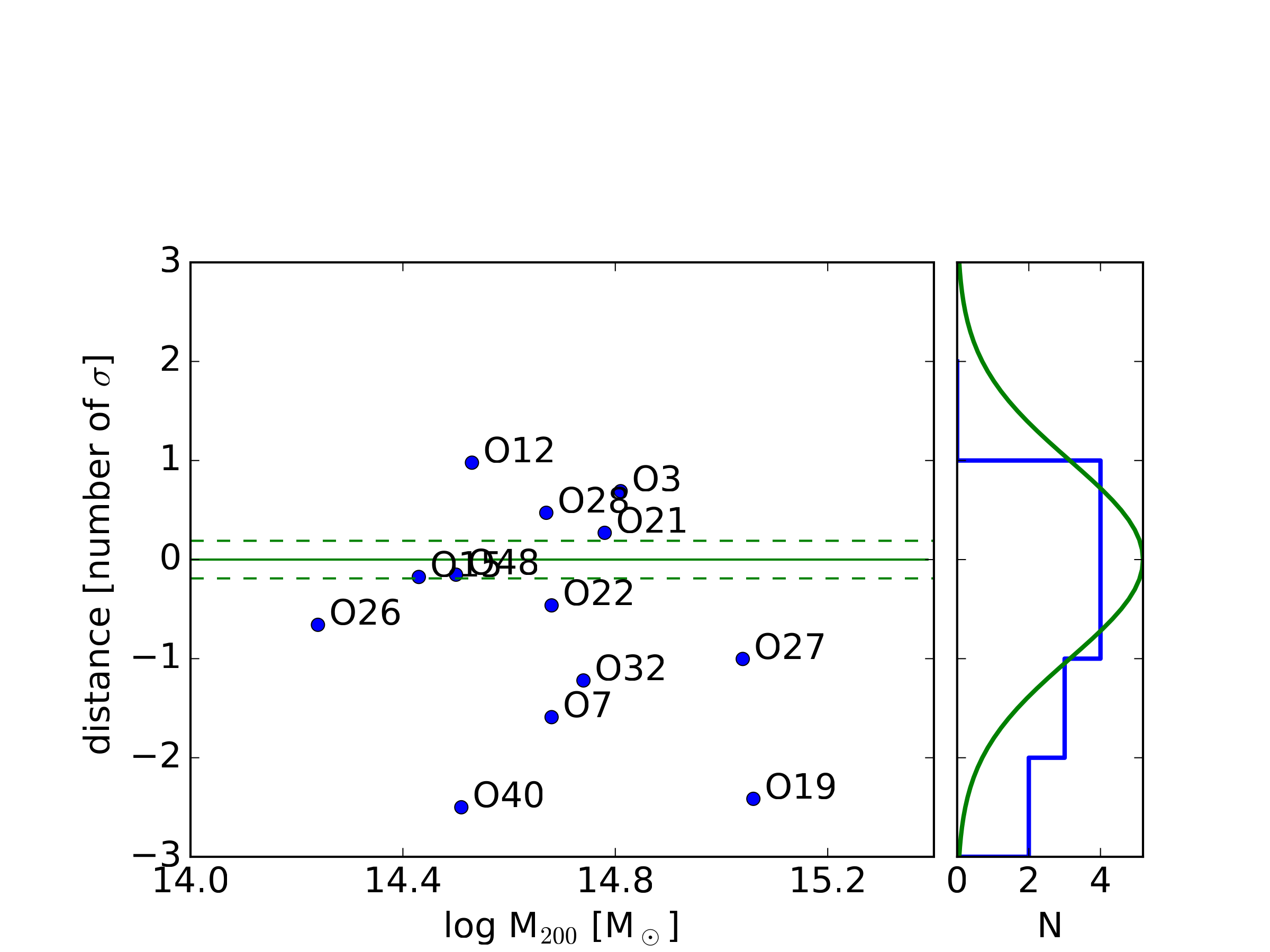}}
\caption{{\it Left:} Minimal signed distance from the bias-corrected mean $Y_{{\rm sph},200}-M$ relation of Andreon (2014). Most distances are negative and in a few cases by $2$ or more $\sigma$, which is unexpected for a sample this small. 
Furthermore, much less than 68\% of the
sample is where it is expected to be, inside the dashed corridor (see text for its derivation). {\it Right:} Marginal distribution. The expected Gaussian distribution is plotted (curve). Manifestly, there is an excess at negative values and a deficit at positive ones.}
\label{fig:Y_M_resid}
\end{figure}

The Planck Collaboration XX (2014) already noticed that the Compton Y vs. mass scaling derived from these 71 clusters would predict more clusters than observed in the Planck survey and hypothesized that their mass estimate was biased by a factor of $1-b$. The bias allows the cluster mass to be larger at fixed Compton Y, or equivalently, the Compton signal to be smaller at fixed mass. The value of the bias has been the subject of ample discussion in the literature, and we use the posterior mean derived in Sec.~3.3 using our data, $1-b=0.72$, for this comparison. The red line shows the Andreon (2014) scaling corrected for the mentioned mass bias. The bias correction moves the scaling toward the gravity-selected sample by construction. Nevertheless, discrepancies remain: first, some clusters, most notably O19, are still offset from it. Second, the distribution
of the clusters in the Compton Y mass plane is uneven:
the signed distance from the bias-corrected Andreon (2014) scaling relation is shown in the left panel of Fig.~\ref{fig:Y_M_resid}. The signed distance is just the minimal distance, times $-1$ for points below the relation, in units of combined errors, derived by approximating the Normal distribution in $Y$ with a Normal distribution in $\log Y$ on the positive side. After bias correction, the distribution of points at $|d|<1\sigma$ is centered around zero, but beyond $1\sigma$ the distribution is very uneven: five clusters have signed distances $<-1\sigma$, and none has $>1\sigma$. There are three clusters with $\lesssim -2$. Instead, fewer than one such cluster is expected (right panel of Fig.\ref{fig:Y_M_resid}). Furthermore, there is a noticeable deficit of clusters at positive values of the minimal distance (Fig.~\ref{fig:Y_M_resid}).
Finally, 68\% of the points are supposed to fall in the green dashed corridor (whose width is given by the quadrature sum of the Andreon 2014 intrinsic scatter and our mean errors on mass and Compton Y) if the bias-corrected ICM-selected scaling were correct. This is clearly not the case. In other words, even after bias correction, the scaling based on the ICM-selected sample does not describe the gravity-selected sample because it is not scattered enough, because it lacks clusters with small  Compton Y for their mass, and because it overpredicts the number of clusters with large Compton Y for their mass. Not correcting for mass bias would increase the disagreement between expectations and observations. Therefore, the Compton Y vs. mass scaling derived from ICM-selected clusters, although said to be corrected for sample selection, is quite different from the one seen using our gravity-selected sample, which by definition does not suffer from ICM-related selection effects.

\begin{figure}
\centerline{\includegraphics[width=9truecm]{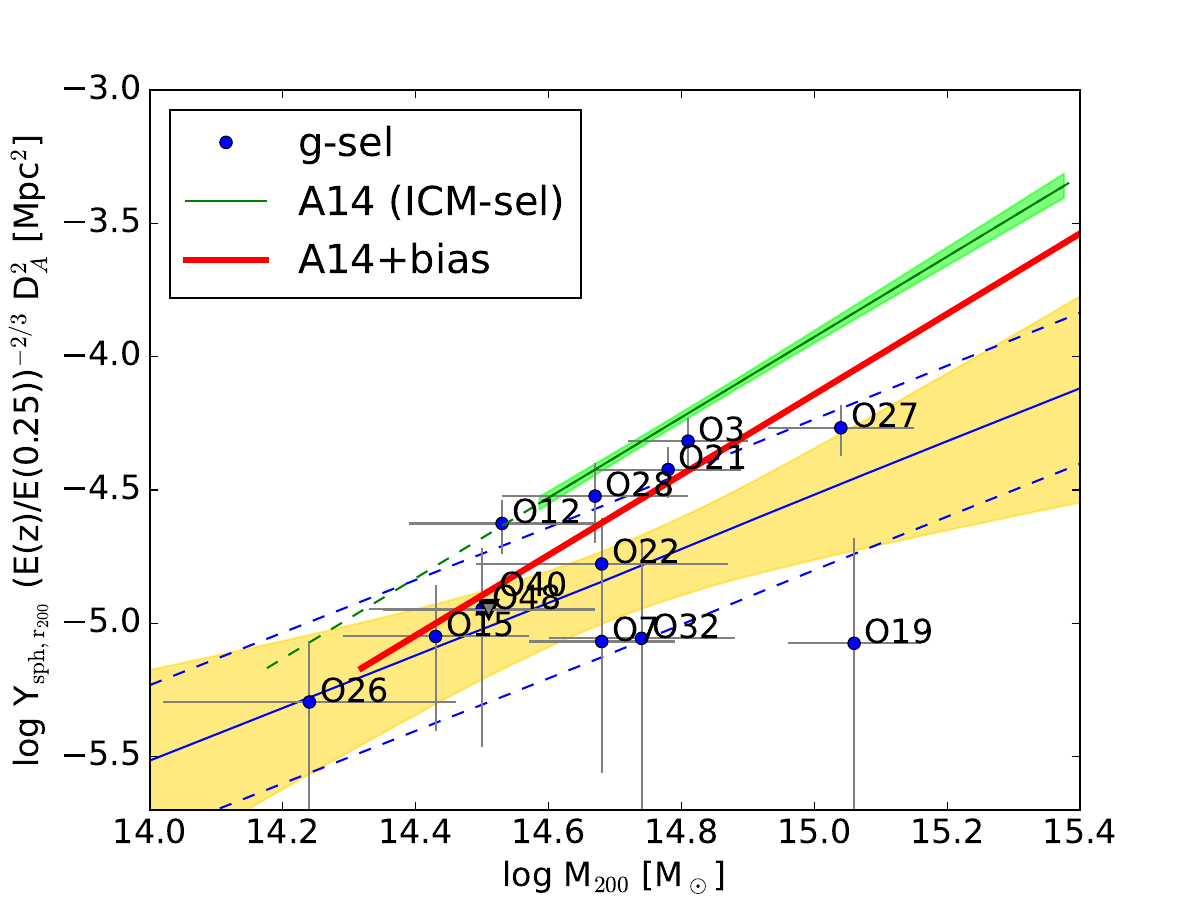}}
\caption{$Y_{{\rm sph},200}-M$ plot 
for our gravity-selected sample alone. The downward pointing triangle indicates the $2\sigma$ upper limit of O40.
The solid line indicates the fitted relation, the shaded (yellow) region its $\pm1\sigma$ uncertainty, whereas the dashed corridor indicates the mean model $\pm$ the median value of the intrinsic scatter. The ICM-based scaling relation is
clearly overstimating the Compton Y -mass relation, as also shown in Fig.~\ref{fig:post}}
\label{fig:Y_M_usalone}
\end{figure}

\subsection{Hypothesis-parsimonious analysis}

Fig.~\ref{fig:Y_M_usalone} shows the Compton Y vs. mass plot for our gravity-selected sample alone. We fit these data assuming a linear relation between $\log$ Compton Y and $\log$ mass, with a Gaussian intrinsic scatter. Errors are modeled as Gaussian in Compton Y and $\log$ mass, which naturally accounts for the upper limit in Compton Y of O40 (i.e., a negative observed value for positively-defined Compton Y, such as the one for O40, is allowed, see Chap. 8.6 of Andreon \& Weaver 2015). As in Andreon (2014), we used uniform priors for the scatter, the intercepts, and the angle (i.e., a Student-t distribution with 1 degree of freedom on the slope, Andreon \& Hurn 2010), but, unlike Andreon (2014) and past literature, we compute the scaling relation at $z=0.25$. Since the Eddington (1913) bias is negligible and has already been corrected for, we do not correct a second time for it (i.e., adopting a non-uniform mass prior, see Sec.~4.1 for a discussion). As discussed in Sec.~2.3, radial covariance is neglected because it was found to be negligible. Posteriors are sampled with a Gibbs sampler (JAGS, Plummer 2010) using a minor editing of the code used for modeling the Compton Y vs. mass relation in Andreon (2014) and distributed in Andreon \& Weaver (2015).
We found: 
\begin{eqnarray}
\log Y_{sph,200}  = (1.0\pm0.6) (\log (M_{200}/M_\odot) -14.788) \nonumber \\
 -4.78\pm0.17  +2/3 (\log (E(z)/E(0.25)) \ ,
\end{eqnarray}
where $E(z)$ is the usual Hubble ratio term with fixed self-similar evolutionary slope of 2/3. Identical parameters are
found leaving the $E(z)$ power coefficient to be free or adopting for it the posterior mean
determined in Andreon (2014).
As evident from Fig.~\ref{fig:Y_M_usalone}, there are no obvious outliers in the relation when gravity-only clusters
are fitted and unsurprisingly
we found the same scaling parameters using a Student-t distribution with 10 degrees of freedom for the intrinsic scatter around the mean relation (this approach makes the determination robust to outliers; see Andreon \& Hurn 2010).
Of course, the absence of outliers for this fitted relation does not imply the absence of outliers from any other scaling relation, in particular Andreon (2014), whether or not is bias-corrected.

\begin{figure}
\centerline{\includegraphics[trim=0 180 270 0,clip,width=4.5truecm]{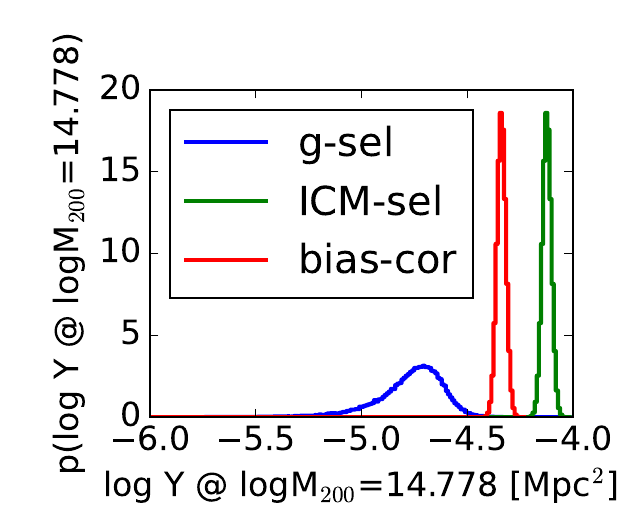}%
\includegraphics[trim=0 180 270 0,clip,width=4.5truecm]{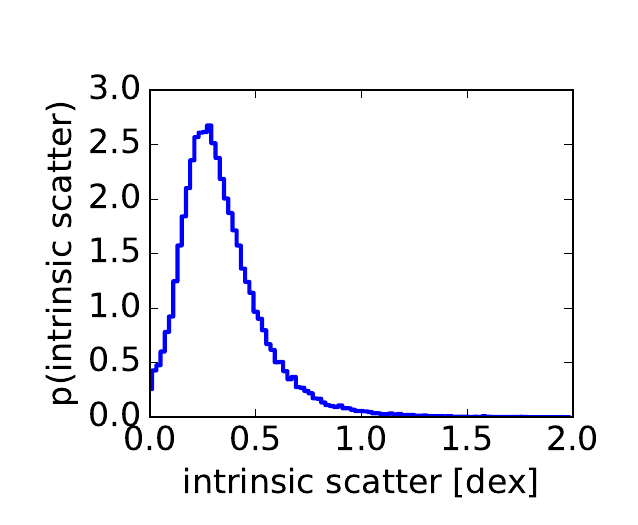}%
}
\caption{{\it Left panel}:  
Posterior distribution of the intercept for the gravity-selected sample and
the ICM-selected sample. Clearly, ICM-selected clusters have a larger Compton Y for their mass.
{\it Right panel}: Posterior distribution of intrinsic scatter for the gravity-selected
sample. Likely, the scatter around the scaling relation is non-negligible, but
better data are needed to precisely determine its value.
}
\label{fig:post}
\end{figure}

Fig.~\ref{fig:Y_M_usalone} also shows the Compton Y vs. mass scaling derived from the ICM-selected sample by Andreon (2014) at $z=0.25$. 
The scaling relation derived from gravity-selected clusters has a lower intercept, as better visible in the left panel of Fig.~\ref{fig:post}, which shows the posterior probability of the intercept (i.e., the mean $\log Y$ value at $\log M_{200}/M_{\odot}= 14.788$ and $z=0.25$, see eq.~1) of our gravity-selected sample and the ICM-selected sample in Andreon (2014). These are manifestly offset each other.

The right panel of Fig.~\ref{fig:post} shows the posterior probability of the intrinsic scatter. We found an intrinsic scatter $\sigma_{intr}=0.30^{0.17}_{-0.13}$ dex suggesting a non-negligible scatter. Although the scatter with a size equal to, or near to, zero has a lower probability than most non-zero values of the scatter, the data still permit it ($p(0)\neq0$). In order to understand if this large uncertainty on the intrinsic scatter arises from the uncertain slope, we re-fit the data assuming the Andreon (2014) slope posterior, $1.51\pm0.07$, as a prior. The value and error of the intrinsic scatter (and of the intercept as well) remained unchanged, indicating that the large uncertainty of the intrinsic scatter is unrelated to the slope uncertainty and is only due to small sample size and data noisiness. In particular, the clusters with small Compton $Y$ values have large $Y$ errors, and reducing these through, for example, follow-up SZ observations, would significantly improve the intrinsic scatter determination.

\begin{figure}
\centerline{\includegraphics[width=9truecm]{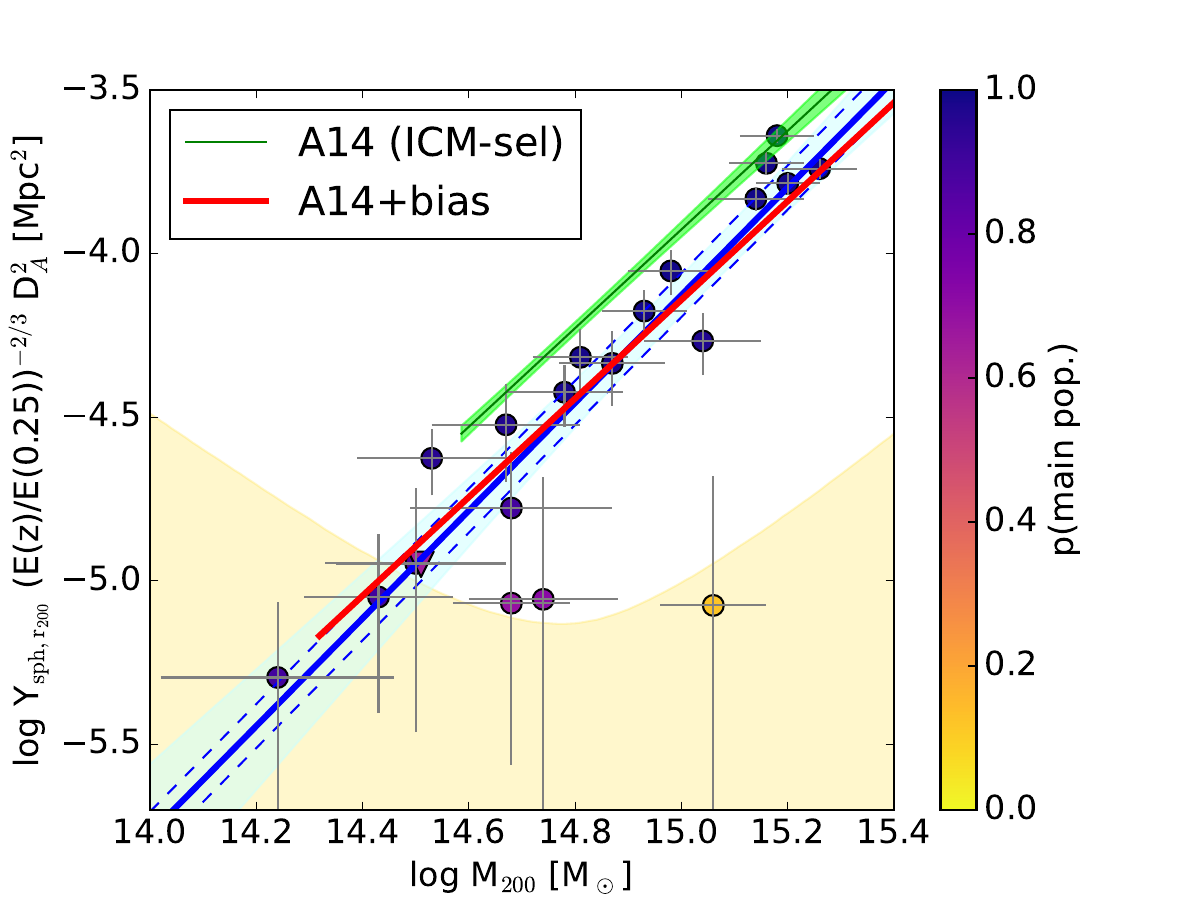}}
\caption{$Y_{{\rm sph},200}-M$ plot 
for gravity and X-ray selected samples of our mixture of regression analysis. The downward pointing triangle indicates the $2\sigma$ upper limit of O40. Points are colour-coded according to probability to belonging to the main population. Shadings indicate the 68\% error on the fit, whereas the dashed corridors indicates  $\pm 1\sigma_{intr}$ around the mean relation of the main population. 
O19 (the orange point) is currently the only secure representative of the population of clusters with a small Compton Y for their mass.
The solid green and red lines indicates the Andreon (2014) relation fitted to the Planck sample before and after mass bias correction. By construction, the mass bias makes the Andreon (2014) relation in agreement with the fit of the main population.}
\label{fig:Y_Mtwopop}
\end{figure}

\subsection{Mixture of regressions}

We now fit our data under the hypothesis that there exist two populations of galaxy clusters, one with a large Compton Y and one with a small one for their mass, and with proportions to be determined from the data, each one obeying a $Y-M$ scaling relation. One of the two populations can be, in principle, composed of zero members. This analysis is risky because we now assume that selection effects for our X-ray-selected sample are negligible in the sense that the population of clusters with large Compton Y does not significantly extend below the region sampled by our data (e.g. there are almost no clusters at $\log (M_{200}/M_\odot)\sim15.4$ and $\log Y\sim -4$ Mpc$^2$) allowing us to neglect accounting for the selection function of our X-ray-selected sample. Of course, at large masses, clusters belonging to the population with small Compton Y at $\log Y\sim -4$ Mpc$^2$ are allowed.

For this analysis, we fit the whole dataset of gravity-selected and X-ray-selected clusters with a mixture of two linear regressions with a Gaussian intrinsic scatter around them, without assuming which points are drawn from which population, but with the strong assumption just mentioned. We took uniform priors for the probability of belonging to the population with smaller Compton Y, for the scatters, the intercepts, and the angles. We sample the posterior with a Gibbs sampler (JAGS, Plummer 2010). Technically, it is just a matter to add to the model of the previous section an indication variable, the membership, following Chap. 7.3 of Andreon \& Weaver (2015). For parameter identifiability, we decide to call ``first" (index 0) the population with smaller Compton Y at fixed mass. For the population with larger Compton Y, that we call the main sequence from now on, we find
\begin{eqnarray}
\log Y_{sph,200}  = (1.7\pm0.2) (\log (M_{200}/M_\odot) -14.788) \nonumber \\
  -4.54\pm0.08 +2/3 (\log (E(z)/E(0.25)) \ 
\end{eqnarray}
with an intrinsic scatter around it of $\sigma_{intr}=0.06^{0.06}_{-0.04}$ dex.

\begin{figure}
\centerline{\includegraphics[trim=0 190 250 0,clip,height=4truecm]{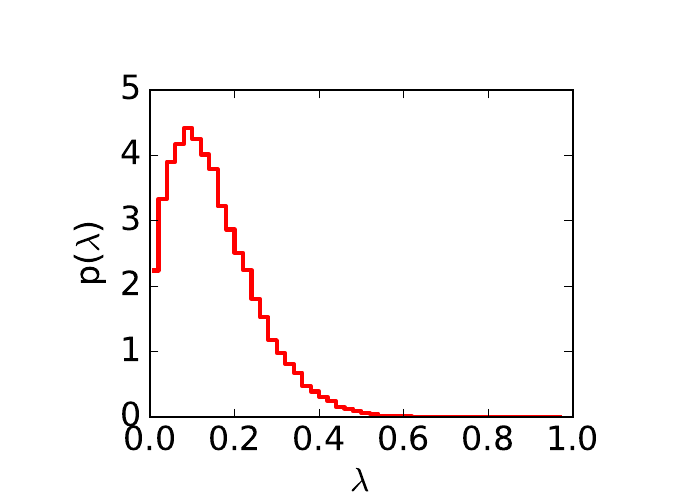}}
\caption{Probability distribution of the fraction of the cluster population with a small Compton Y for their mass. The fraction is non-zero, but better data are needed to precisely determine its value.}
\label{fig:plambda}
\end{figure}

Most clusters belong to the main sequence (in Fig.~\ref{fig:Y_Mtwopop} $p$ is $\sim1$, i.e., most points are blue), and indeed, the population of clusters with small Compton Y for their mass only composes $13^{+10}_{-8}$\% of the whole population (this is one of the fitted parameters,  $\lambda$, whose posterior
probability distribution is depicted in Fig.~\ref{fig:plambda}). O19 is currently the only secure representative of this population,
which already stand out as an outlier in our empirical analysis.
The scaling relation of the the population of clusters with small Compton Y for their mass is largely undetermined, except to have a low intercept and, as mentioned, being poorly populated.

\begin{figure}
\centerline{\includegraphics[trim=100 0 0 0,clip,width=9truecm]{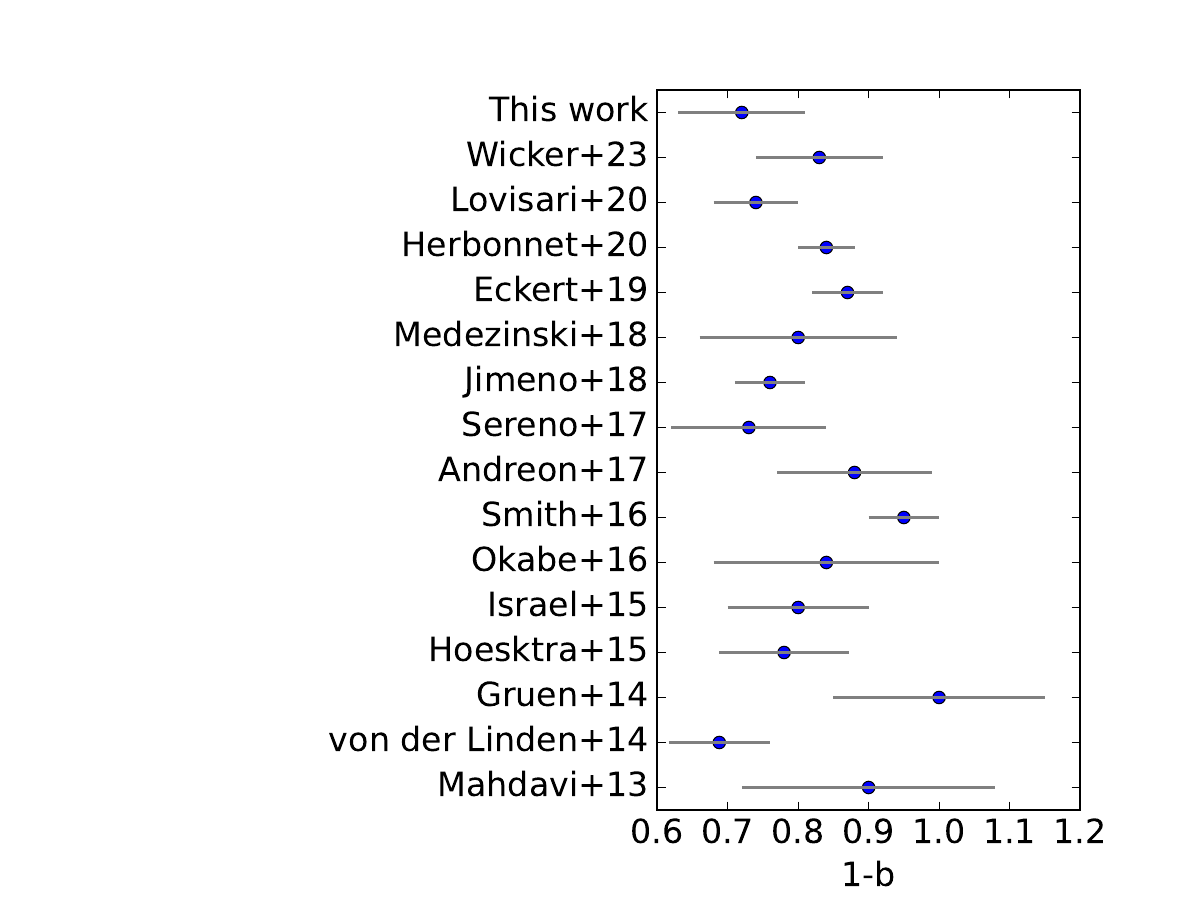}}
\caption{Mass bias $1-b$ derived in our and other works.}
\label{fig:bias}
\end{figure}

The main sequence (blue solid line in Fig.\ref{fig:Y_Mtwopop}) is offset compared to the Andreon (2014) fit to ICM-selected samples (green solid line in the same Figure). The horizontal offset, which can be easily derived as the difference between the intercepts divided by the slope, is found to be $\log (1-b) =-0.15\pm0.05$ or approximately $1-b=0.72\pm 0.09$ (posterior mean and standard deviation), that is the value adopted in previous sections. The error on the bias is almost entirely due to the large uncertainty of the intercept of the scaling relation of the gravity-selected sample and has a negligible dependence on the slope. Fig.\ref{fig:bias} shows that our determination of the mass bias is in line with those in the literature
(Mahdavi et al. 2013, von der Linden et al. 2014, Gruen et al. 2014, Hoekstra et al. 2015, Israel et al. 2015, Okabe \& Smith 2016, Smith et al. 2016, Sereno et al. 2017, Andron et al. 2017b, Jimeno et al. 2018, Medezinski et al. 2018, Eckert et al. 2019, Herbonnet et al. 2020, Wicker et al. 2023). The mass bias is usually understood as 
something affecting the mass determination. However, the Compton Y parameter (or the observable used in other works) is aperture-dependent and the aperture is set by the mass. This implies
that a smaller mass bias suffices to reconcile (the two scaling relations in our case) once the aperture used for measurements is consistent with what it is said to be. The proper determination of the bias, accounting for the aperture covariance, is left to a future work.

The fit describing the main sequence is quite different from the one derived in our hypothesis-parsimonious fit, being steeper and with lower scatter. The tilted slope and the reduced scatter are driven by the added X-ray selected sample with its large values of Compton Y and reduced scatter plus our modelling that assumes that at large masses there are no other main sequence clusters at $\log Y\sim -4$ Mpc$^2$. The latter assumption is unsure because there are no previous analyses, to our best knowledge, of non-ICM selected samples which exclude the existence of clusters at this position of the diagram. Determining whether the scatter and slope determination of our risky analysis are biased awaits a shear catalog with much larger sky coverage, enabling the sampling of very massive, and therefore rare, clusters. Intercept and mass bias are instead already robust with the current sample, and consistent with what found in our hypothesis-parsimonious fit, because the current gravity-selected sample is large enough and has the right mass range for these determinations.

\subsection{Summary of the three analysis}

We performed three analyses, each with varying levels of assumptions.

The empirical approach shows that the Compton Y vs. mass scaling relation derived from an ICM-selected sample, corrected for selection effects and therefore intended to represent the entire cluster population, has a too-large intercept, too-low scatter, and a lack of clusters that have a small Compton Y for their mass compared to our gravity-selected sample. The assumption of a mass bias of $1-b=0.72$ reduces the disagreement, but the bias alone is insufficient to fully reconcile the ICM-selected scaling with the gravity-selected data due to insufficient scatter, overprediction of clusters with large Compton Y for their mass, and underprediction of clusters with small Compton Y for their mass.

The hypothesis-parsimonious analysis, which assumes a single population of clusters and relies solely on the gravity-selected sample, yielded similar results: the intercept of the scaling relation for ICM-selected samples after accounting for sample selection and whether or not are mass-bias corrected, is larger than that derived from our gravity-selected sample.

The analysis with a mixture of regression posits the existence of two distinct populations of clusters, each obeying its own scaling relation, with proportions inferred from the data (including the possibility of one population having zero members), and assumes that the ICM-selected sample used at very large masses is not significantly biased. The results again suggest a larger intercept for ICM-selected samples and imply a mass bias of $1-b=0.72\pm0.09$. Additionally, a second population of clusters with small Compton Y for their mass assumed nonexistent in the analysis of ICM-selected samples, is identified, though it constitutes only a numerically minor component. O19 appears to be the most prominent member of this second population.
These conclusions are robust to the assumption underlying this mixture of regression analysis. However, the determined slope and intrinsic scatter of the main cluster sequence could be biased.

\begin{figure}
\centerline{\includegraphics[width=9truecm]{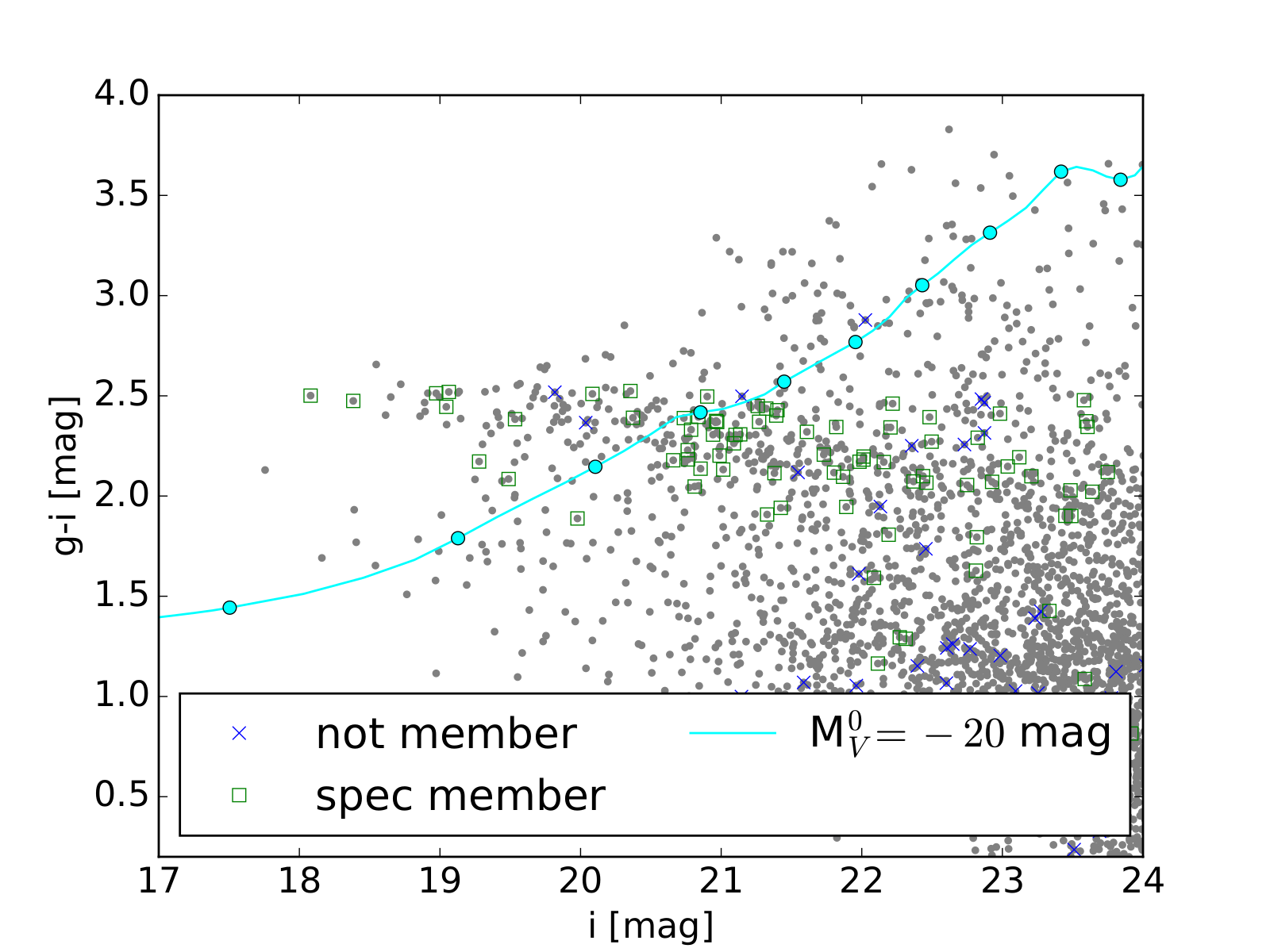}} 
\caption{Color-magnitude plots of all galaxies within $r_{200}$ of the O19 center.
The solid line indicates the intercept of the color-magnitude relation at different redshifts (values multiple of 0.1 are marked with circles starting from $z=0.1$). 
Basically, massive galaxies are in the top left part of the diagram delimited 
by the cyan curve and massive clusters  
can be identified as a red sequence extending to the left of the curve.
Only one red sequence is present, at the color appropriate for the O19 redshift. 
A twice richer cluster is need to make O19 a member of the main sequence, while
the figure shows the lack of other (contaminating) similary rich, and therefore massive, 
clusters along the line of sight. 
}
\label{fig:O19CM}
\end{figure}

\subsection{O19}

O19 is the object with the largest distance from the Compton Y vs. mass relations based on ICM-selected samples (visible in the bottom-right of Fig.~\ref{fig:Y_M}, \ref{fig:Y_M_resid}, and \ref{fig:Y_Mtwopop}). For this reason, it warrants a closer examination.

First, O19's outlier status is not due to the presence of a bright radio source, or millimeter galaxy, filling the SZ decrement, as the cluster is also an extreme outlier in the $L_X-M$ scaling relation in the X-ray domain (Li et al., in prep). In X-ray, unflagged point sources boost the cluster flux, rather than reducing it. As additional precaution, we examined available radio and millimeter archive data, including VLA, ASKAP, TGSS, ACT (220 GHz), and SCUBA-2 observations. No sources were found within 2 arcminutes of the cluster, down to the detection limits of these observations.

We next assess the robustness of the weak-lensing mass estimate, focusing on whether it could be erroneous. The mass estimate is unlikely to be affected by issues with the HSC shear data, as a consistent tangential shear profile is obtained using the Kilo-Degree Survey (KiDS) shape catalog (Kuijken et al. 2019, Wright et al. 2020, Hildebrandt et al. 2021, Giblin et al. 2021). An identical mass is obtained when shifting the cluster center from the coordinates in Oguri et al. (2021) to the brightest cluster galaxy. This is expected, given that the shift is 130 kpc while the innermost radial bin considered is 400 kpc.
The richness-based mass of O19, derived exactly as those of the clusters in Andreon et al. (2025) from the same source of data using the same code, is $\log M/M_{\odot} = 14.93\pm0.16$, in line with the weak-lensing estimate and indeed the most massive cluster among our 
gravity-selected sample (Andreon et al., in prep) according to the richness-based estimates.

For the cluster to align with its Compton Y value, its mass would need to be overestimated by a factor of three (or by a factor of five to lie exactly on the scaling relation). This would require another outlier cluster, with at least twice the mass that would put O19 on the scaling ($\log M/M_{\odot} = 14.9$) but with Compton Y emission corresponding to a cluster of mass $\log M/M_{\odot} < 14.2$, along the line of sight. Thus, to avoid an oulier we need
to postuate the existence of another outlier (on the line of sight), which is inherently nonsensical. 

Furthermore, this postulated outlier should be nearly devoided of galaxies. In fact,
optical data (e.g., the color-magnitude diagram in Fig.~\ref{fig:O19CM}) definitively rule out the presence of another massive cluster along the line of sight at $\delta z \gg 0.02$. A second red sequence, indicative of another cluster, would be clearly visible (e.g. Andreon \& Berge 2012), even if the additional cluster had a mass significantly smaller than O19's (and it should be twice as much to make O19 a member of the main sequence). 

Let's now consider the remaining explanation, a highly elongated O19 cluster with an axis ratio of about 3
where two-thirds of the cluster would be ``dark" in Compton Y, or a configuration involving two subclusters,
one of which is twice as massive as O19 yet nearly Compton Y dark. Spectroscopic data rule out these
scenarios unless the hypothetical outlier is also nearly galaxy-free. First, of the 16 available SDSS
redshifts for galaxies in the field, 12 fall within a single peak, with a velocity dispersion of  $\approx
500$ km/s. Second, the deep spectroscopic survey by Griffiths et al. (2018), with 89 members, reveals a
single peak with a velocity dispersion of 600 km/s and no evidence of additional peaks. This rules out the
presence of secondary, or tertiary, overdensities containing a significant fraction of O19's mass. The
spectroscopic data also confirm that there are no clusters more massive than O19 within $\delta z > 0.02$.

To summarize, the only way O19 can not be an outlier is to postulate the existence of another outlier,
which is is inherently nonsensical, and it should be nearly
devoided of galaxies. We are therefore obliged to conclude that O19 is the member of a new population of galaxy clusters
with a small Compton Y for their mass.
Higher quality SZ data and X-ray data would
be very useful to characterize O19, for example to check whether it is a merging in progress.

\section{Discussion}

\begin{figure}
\centerline{\includegraphics[width=9truecm]{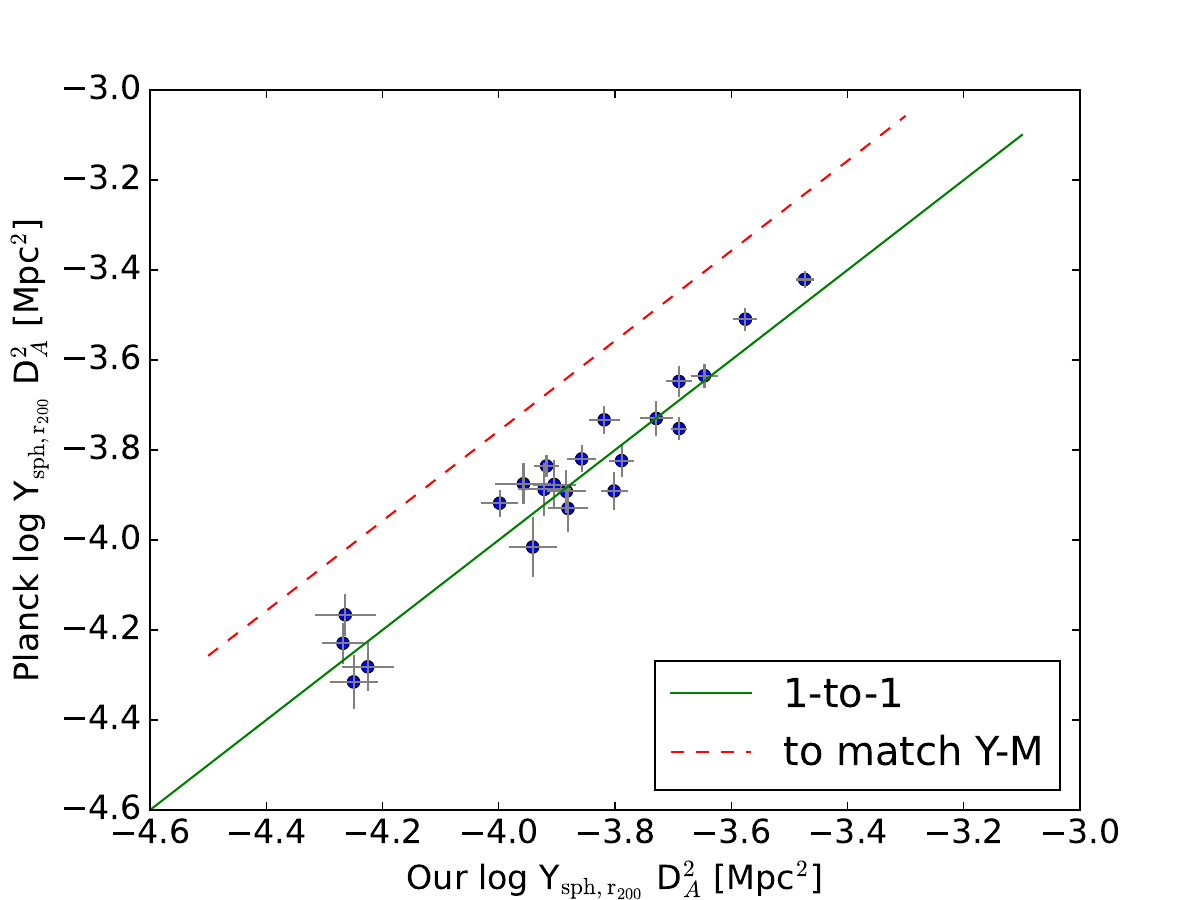}}
\caption{Comparison between our and Planck photometry for Planck clusters in the ACT footprint. 
The disagrement needed to match the Y-M scaling relations of ICM- and gravity- selected samples is
indicated as red dashed line and is clearly incosistent with the data, that instead support no bias in photometry
across sources of data.}
\label{fig:photom_bias}
\end{figure}

\subsection{Compton Y Systematics}

To check whether the offset between gravity- and ICM- selected scaling relation is due to a bias in Compton Y photometry
across sources of Compton Y data,  
we now measure the Compton Y of  the 22 Planck clusters inside the ACT footprint and in the same redshift range
studied in this work and compare them to the Planck photometry. We use Planck Coll. (2014) $r_{200}$ aperture, estimated from $M_{500}$, and ACT maps, as done for the
gravity-selected sample. Fig.~\ref{fig:photom_bias} shows the good agreement between Planck Coll. (2014) and our (ACT)
photometry, unsurprising given that the ACT Compton map incorporates the Planck photometric data. In particular, the data clearly
discard the bias needed to match gravity- and ICM- selected clusters, indicated with a dashed red line.

\subsection{Weak-lensing Systematics}

Possible systematics in the galaxy shape measurement from the first-year HSC shear catalog were reviewed by Mandelbaum et al. (2018), who showed that additive and multiplicative biases are all within the requirements. 
Cluster triaxiality may lead to a scatter, which is accounted for in our mass estimation (Sec~2.1).
The inclusion of foreground and cluster-member galaxies in the sample of lensed galaxies can introduce a dilution of the weak lensing signal and therefore an underestimate of the cluster mass. Medezinski (2018) showed that the adoption of the P-cut (Oguri 2014) selection method used in this analysis allows to  effectively minimize this contamination (see also Hamana et al. 2020).   
As discussed in Euclid collaboration: Giocoli et al. (2024), modelling both the mass and concentration with a NFW halo profile can introduce an underestimate by $\sim$ 5-10\%  in   clusters masses estimated by weak lensing, that reduces to $<$ 5\% if the concentration is fixed, in agreement with earlier works (e.g. Becker \& Kravtsov 2011). These effects are way too small to affect our results and, notably, underestimating masses would amplify discrepancies with the ICM-selected sample. As a way of reference,
Chiu et al. (2024) utilize the latest release of the shear catalog that we employ to analyze a gravity-selected cluster sample. They fit the shear data using a spherical NFW profile, excising the cluster center in a manner similar to our approach. For their cosmological analysis, they adopt a simulation-based weak-lensing mass bias of $M_{WL}/M=0.99\pm0.05$ (compared to our value of 1) and a scatter of $24\pm4$\% (versus our value of 20\%).

\subsection{Modeling the S/N selection in analysis?}

Our analysis employs a uniform prior for the Eddington-corrected masses. A devil's advocate might argue that this approach overlooks the signal-to-noise (S/N) cut we applied and that we should instead explicitly model the S/N selection function.

The Eddington correction accounts for the non-uniform prior distribution of objects in the parameter of interest, in our case, the steep mass function. Researchers may choose to study objects above a certain mass or S/N threshold instead of randomly selecting objects. However, this decision does not alter the fundamental physical effect that any selected sample is more likely to include lower-mass objects due to observational scatter. The Eddington bias remains intrinsic to the process and independent of researcher-imposed selection criteria. Such criteria are typically applied after the bias has already occurred, and it is implausible for nature to ``anticipate" the thresholds imposed by researchers.

\begin{figure}
\centerline{\includegraphics[trim=0 0 0 0,clip,width=9truecm]{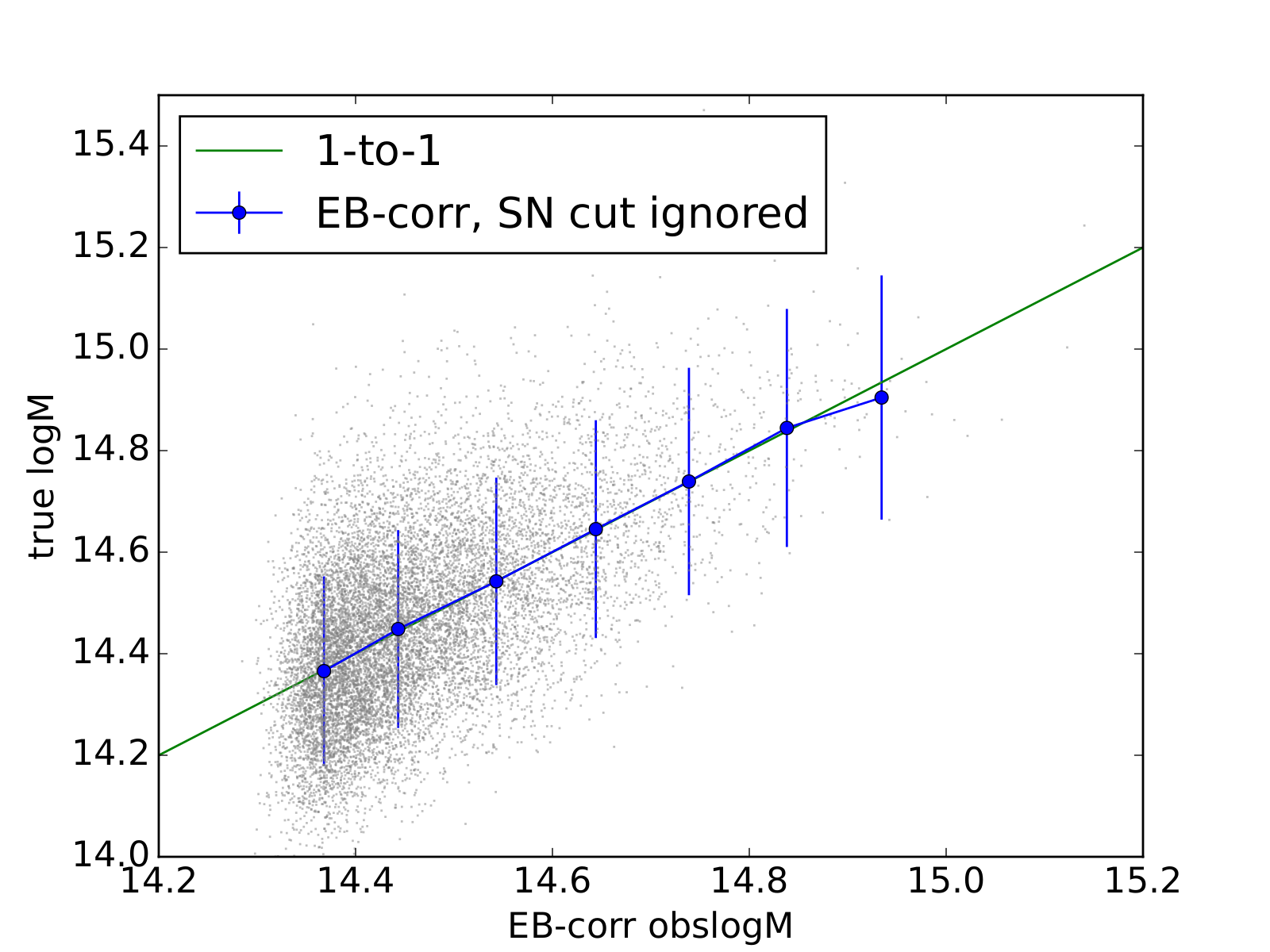}}
\caption{Lack of effect of a $S/N>5$ selection on the estimated mass: the Eddington-corrected value, derived ignoring
the $S/N$ selection, is unbiased compared to the true value. The shown errorbars indicate the average error of masses used
in our simulation, tailored around our real data. The ``bias" of the $S/N$ selection is, if any, an order of magnitude smaller than mass error.}
\label{fig:effect_of_cut}
\end{figure}

To clarify this concept, we provide a numerical example. Cluster masses are drawn from the Tinker et al. (2008) mass function (chosen for specificity, though the result is independent of the chosen function). These masses are "observed" by applying scatter consistent with realistic, mass-dependent measurement uncertainties. Each cluster is assigned an observed S/N value based on the observed mass, with additional scatter (the results are insensitive to the magnitude of this scatter). Using the prior (the Tinker et al. 2008 mass function), we compute the Eddington-corrected masses. As expected, the corrected masses show no bias or offset compared to the true values.

Next, we impose an S/N selection criterion, requiring $S/N > 5$. To explore the implications, we adopt a lower threshold than in our analysis to simulate more relaxed future selection criteria. Importantly, the choice of threshold does not affect the conclusions. Despite this S/N selection, the Eddington-corrected masses remain unbiased and aligned with the true values, as illustrated in Fig.~\ref{fig:effect_of_cut}. This demonstrates that S/N selection does not need to be explicitly accounted for in the analysis.

Even in the unlikely scenario where both our theoretical framework and simulation methodology were flawed, the resulting mass estimates would be underestimated by only a fraction of the Eddington correction. Since the Eddington correction itself is negligible in our analysis, this hypothetical error would remain inconsequential. Notably, underestimating masses would amplify discrepancies with the ICM-selected sample, further emphasizing the robustness of our findings.

\begin{figure}
\centerline{\includegraphics[width=9truecm]{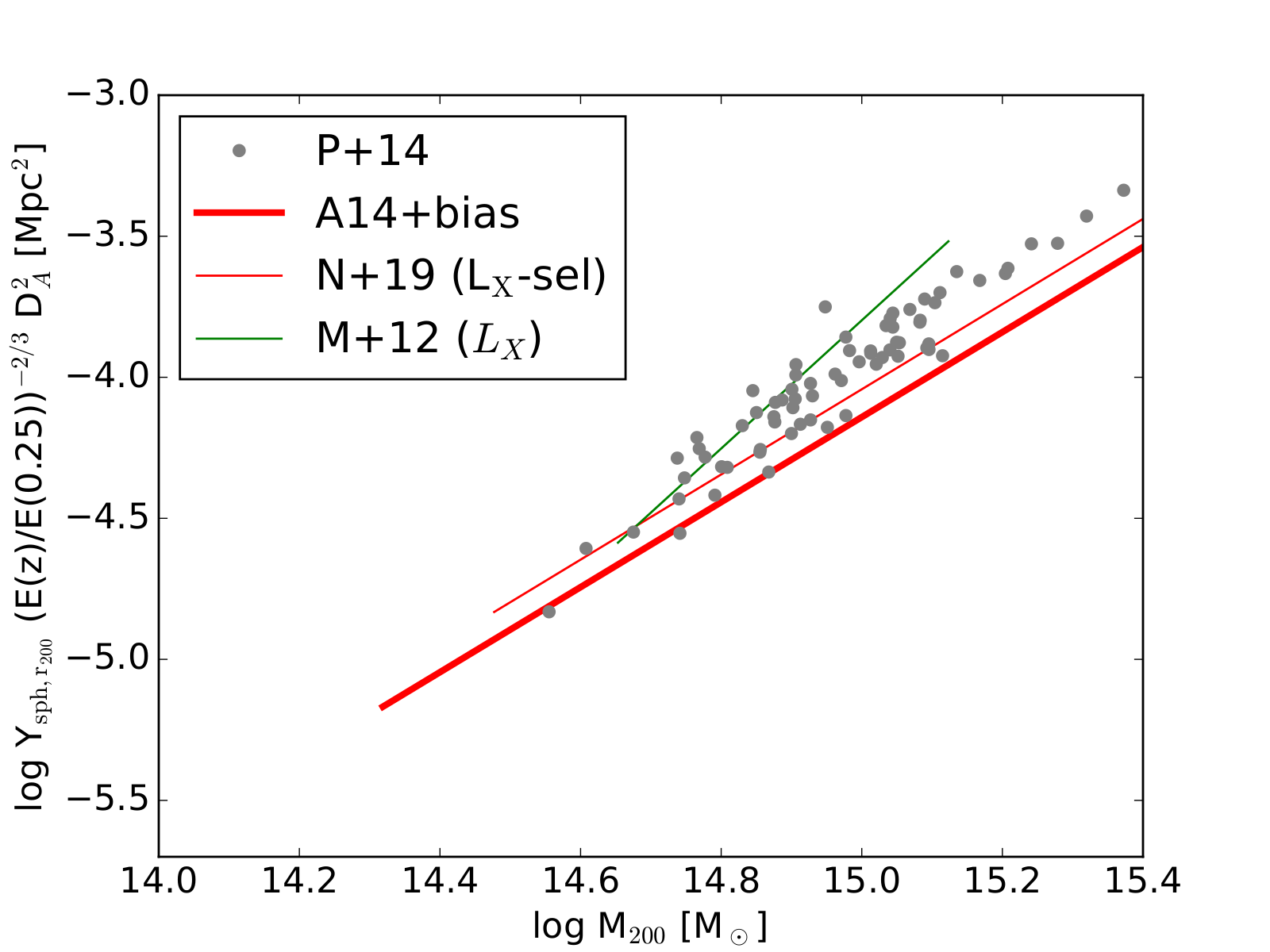}}
\caption{Literature $Y_{{\rm sph},200}-M$ scaling relations. 
Gray points are clusters from Planck Collaboration XX (2014).
Literature scaling
relations overestimate the Compton Y signal at fixed mass compared to the bias-corrected Andreon (2014) scaling relation, which, in turn,
overestimates the scaling relation of the gravity-selected sample, making our conclusions robust to the choice of the comparison sample.
}
\label{fig:Y_M_lit}
\end{figure}

\subsection{Other Compton Y-mass relations}

A devil's advocate might argue that our results are driven by the comparison, ICM-selected, sample having a too large Compton Y for its mass (for unknown reasons).
Let's therefore consider two other literature scaling relations, Marrone et al. (2012) and 
Nagarajan et al. (2019), both estimated at $z=0.25$ from their extrapolated (by the authors) value at $z=0$ assuming self-similar
evolution. 
The literature relations, shown in Fig.~\ref{fig:Y_M_lit}, have a larger Compton Y signal at fixed mass than the bias-corrected
Andreon (2014) scaling relation, taken as comparison sample in previous sections, which in turn overestimate the Compton Y signal and does not describe the scaling of gravity-selected
clusters, confirming the robustness of our results.
These two
scaling relations suffer of limitations, however, in addition to the mentioned wrong assumed evolutionary correction. Marrone et al. (2012)
is clearly too steep compared to the Planck and our data. Marrone et al. (2012) does not account for Eddington bias, for sample selection, and uses a wrong prior (see Sec.~4.3). 
Nagarajan et al. (2019) uses weak-lensing masses and therefore their closiness to the bias-corrected Andreon (2014) relations is
expected, although we note that it has large errors, being 16 times less constrained than the latter. The remaining offset could be related to
fact 
that their Compton $Y$ measurements were performed at radii corresponding to the maximum likelihood estimate of the weak-lensing mass. This approach, which does not correct for Eddington bias the aperture used for the photometry, leads to an overestimation of the Compton $Y$ flux. Applying it would likely bring their determination close to our main sequence and to the bias-corrected Andreon (2014) scaling relation.

\subsection{Orlowski-Scherer et al. 2021}

Orlowski-Scherer et al. (2021) modeled the population of clusters at fixed richness using a double Gaussian distribution, with the second component accounting for objects with a small Compton Y signal for their richness. They found that the relative proportion of large Compton Y to small Compton Y clusters is 57\% at $z \sim 1$, a value significantly larger than the $13^{+10}_{-8}$\% proportion we identified at $z \sim 0.25$. This discrepancy may reflect evolutionary effects, given the 5 Gyr look-back time difference between the two samples' mean redshifts. Alternatively, it could stem from contamination in the Orlowski-Scherer et al. (2021) sample. Their richness estimates, from Gonzalez et al. (2019), rely on a color selection that shifts by 0.1 mag per $\Delta z = 0.1$, while their photometric data often have uncertainties of this magnitude. This poor redshift resolution leads to significant contamination from line-of-sight groups, which spuriously inflate richness while having minimal impact on Compton $Y$.
A less contaminated 
sample will be critical for disentangling two competing explanations: whether the $z \sim 1$ clusters with small Compton Y identified by Orlowski-Scherer et al. (2021) represent a genuine population of unvirialized objects with reduced intracluster medium (a physical cluster property) or whether their findings are primarily due to observational challenges, specifically line-of-sight projection effects that artificially inflate richness measurements.

\section{Conclusions}

We study the scaling relation between the Compton Y parameter and the mass of a gravity-selected sample of galaxy clusters, i.e., a sample selected based on the effect of gravity on the shapes of background galaxies. Unlike ICM-selected samples, gravity-selected samples have the advantage of not requiring corrections for selection biases in the intracluster medium (ICM) property being investigated, such as Compton Y, because that property is not used in the sample selection. 

We considered a gravity-selected sample of 13 clusters with weak-lensing signal-to-noise ratios greater than 7 and redshifts in the range $0.1 < z < 0.4$, taken from the peak catalog of shear-selected clusters in Oguri et al. (2021). We determined spectroscopic redshifts using SDSS spectroscopy, derived tangential radial profiles from HSC shear data, and calculated cluster masses by fitting a Navarro, Frenk, \& White (1997) profile to the tangential shear profiles. Compton Y measurements at the overdensity $\Delta=200$ were obtained from ACT maps.

We complemented it with an X-ray selected sample of 8 clusters in the same redshift range and with 
weak-lensing signal-to-noise ratios greater than 7. We compared these two samples with the 71 clusters
in the Planck collaboration XX (2014) that are either X-ray or SZ-selected (i.e., ICM-selected), with
$Y_X$-based masses and with selection effects corrected for and we show that our results results are robust
to the choice of the comparison sample.

We performed three analyses of the Compton Y vs mass scaling, each involving different assumptions, and all agree that
the scaling relation between Compton Y and mass of the gravity-selected sample is not well described by the usual scaling derived from ICM-selected samples because the analysis of those samples tend to underestimate the diversity of the cluster population. They miss clusters with a small Compton Y for their mass and underestimate cluster masses. According to our analysis, this mass underestimation is quantified by a bias factor of $(1-b) = 0.72 \pm 0.09$. Even after correcting the scaling relation from ICM-selected samples and hydrostatic masses for this bias,  we uncover tantalizing indication for a seconday population with small Compton Y which comprises  $13^{+10}_{-8}$\% of the whole population. O19 is currently the only secure representative of this population, with
a Compton Y signal smaller than what would be expected for a cluster 
with a mass one-third of the mass we estimate. To avoid classifying O19 as an outlier, we would have to assume the existence of an
outlier along the line of sight, which is inherently nonsensical. Furthermore, such a hypothetical outlier should be almost devoid of galaxies because undetectable in optical and spectroscopic data. We are therefore obliged to conclude that O19 is the member of a new population of galaxy clusters with a small Compton Y for their mass.

To summarize, the scaling relation between the Compton Y parameter and mass is not well described by the usual scaling derived from ICM-selected samples. Even after accounting for selection effects and mass biases, analyses based on these samples exhibit an excessively large intercept, overly low scatter, and a lack of clusters of the second popultation of clusters 
with a small Compton Y for their given mass.

\begin{acknowledgements}
We thank Nobuhiro Okabe for providing us sky coordinates of their cluster sample,
and Raphael Wicker for the values plotted in Fig.~\ref{fig:bias}. SA thanks Marco De Petris, Jack Orlowski-Scherer,  Nobuhiro Okabe,
Raphael Wicker, Sebastain Grandis, Masamune Oguri, and I-Non Chiu for comments on the paper. SA also thanks Rogier Windhorst, Adi Zitrin, Brenda Frye, and Rachel Honor for discussion about O19.
SA acknowledges INAF grant  
``Characterizing the newly  discovered clusters of low surface  brightness" and PRIN-MIUR grant
20228B938N ``Mass and selection biases of galaxy clusters: a multi-probe approach", the latter funded by the European 
Union Next generation EU, Mission 4 Component 1  CUP C53D2300092 0006.
\end{acknowledgements}

\vspace{5mm}
\facilities{Subaru Telescope (Hyper-Suprime Camera), ACT, SDSS}

{}

\appendix{}
\section{Griffiths et al. 2018}

O19, also known as CLIO (Griffiths et al. 2018), has been described by the authors as a unique cluster due to its unusually compact and concentrated nature. This conclusion, however, is based solely on galaxy number counts and not on mass determinations. Our tangential shear profile data, bounded to $r > 350$ kpc for safety reasons, prevents us from reliably measuring the cluster concentration with the required precision. \

It is important to emphasize that the mass estimates for O19 provided by Griffiths et al. (2018) should be regarded as very preliminary, as noted by the authors and therefore cannot be compared to our mass estimate. For simplicity, the authors employed an outdated velocity dispersion estimator, known to be biased under ideal conditions (Andreon et al. 2008), and even more so given that they sampled only the cluster's central region rather than the spatial scale of interest ($r_{200}$). Furthermore, their analysis neglected the surface term correction (The \& White 1986) and assumed dynamical equilibrium.
As an alternative mass estimate, Griffiths et al. (2018) extrapolated the weak and strong lensing mass measured at $r < r_{200}/10$ (=90 kpc) to $r_{200}$ ($\sim 1$ Mpc). 

\label{lastpage}

\end{document}